\documentclass[aps,prb,twocolumn,nofootinbib,superscriptaddress]{revtex4-2}
\usepackage[dvipdfmx]{graphicx}
\usepackage{dcolumn}
\usepackage{times}
\usepackage{color}
\usepackage{bm}
\usepackage{braket}
\usepackage{physics}
\usepackage[version=4]{mhchem}
\usepackage{amsmath, amssymb}
\usepackage[colorlinks=true, citecolor=blue]{hyperref}

\usepackage{multirow}
\newcommand{\nn}{\nonumber}

\hypersetup{
setpagesize=false,
 bookmarksnumbered=true,%
 bookmarksopen=true,%
 colorlinks=true,%
 linkcolor=black,
 citecolor=blue,
}

\begin{document}
\preprint{APS/123-QED}
\title{
Theory of Andreev and shot noise spectroscopy for topological superconductors \\ probed by $s$-wave superconducting tips
}

\author{Jushin Tei}
\email{tei@blade.mp.es.osaka-u.ac.jp}
\affiliation{Department of Materials Engineering Science, The University of Osaka, Toyonaka, Osaka 560-8531, Japan}
\author{Ryo Hanai}
\affiliation{Department of Physics, Institute of Science Tokyo, Meguro, Tokyo 152-8551, Japan}
\author{Satoshi Fujimoto}
\affiliation{Department of Materials Engineering Science, The University of Osaka, Toyonaka, Osaka 560-8531, Japan}
\affiliation{Center for Quantum Information and Quantum Biology, The University of Osaka, Toyonaka, Osaka 560-8531, Japan}
\affiliation{Center for Spintronics Research Network, Graduate School of Engineering Science, The University of Osaka, Toyonaka, Osaka 560-8531, Japan}
\author{Takeshi Mizushima}
\affiliation{Department of Materials Engineering Science, The University of Osaka, Toyonaka, Osaka 560-8531, Japan}
\date{\today}

\begin{abstract}
    Scanning tunneling microscopy (STM) and spectroscopy (STS) with $s$-wave superconducting tips has been widely applied to probe exotic superconductors, but its potential for investigating topological superconductors remains unclear.
    In junctions between an $s$-wave superconductor and a topological superconductor, the dominant tunneling process is Andreev reflection, in which Cooper pairs from the $s$-wave superconductor tunnel as particle--hole excitations into the surface state of the topological superconductor.
    In this work, we theoretically investigate the fundamental properties of Andreev and shot noise spectroscopy on topological superconductors, focusing on the $dI/dV$ characteristics and current noise.
    We develop a real-time description of an effective tunneling action incorporating Andreev reflection processes in the Keldysh formalism and derive analytical expressions for the Andreev reflection current and the associated current noise.
    Furthermore, we perform numerical simulations for representative topological superconductors and provide a catalog of $dI/dV$ spectra and the Fano factor. 
    Our results establish guidelines for probing topological superconductivity using STM with $s$-wave superconducting tips, and provide theoretical benchmarks for future STS experiments.
\end{abstract}

\maketitle

\section{Introduction}
\begin{figure}[t]
    \centering
    \includegraphics[width=\linewidth]{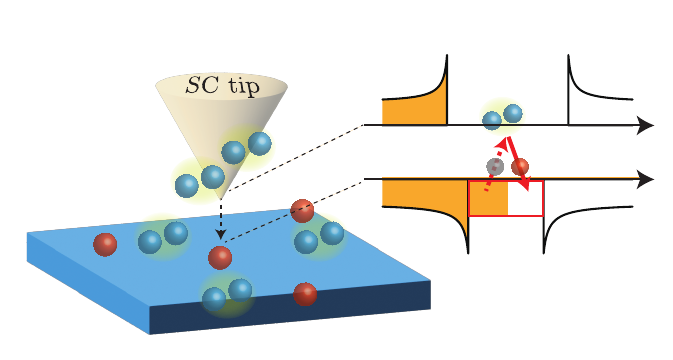} 
    \caption{Schematic illustration of STM measurements with an $s$-wave superconducting tip for a topological superconductor.
    In the low-bias regime, the tunneling current is dominated by Andreev reflection processes, where Cooper pairs from the $s$-wave superconductor tunnel into the particles and holes within the surface states of the topological superconductor.}
\end{figure}

Scanning tunneling microscopy (STM) and spectroscopy (STS) have become an indispensable tool in the study of superconductivity~\cite{hess1989scanning,fischer2007scanning,hoffman2011spectroscopic}.
By directly probing the local density of states (DOS) with atomic resolution, STM has enabled the visualization of a variety of fundamental phenomena, ranging from quasiparticle excitations in the superconducting energy gap to vortex-core states~\cite{hess1989scanning,hess1990vortex,maggio1995direct}, impurity-induced bound states~\cite{yazdani1997probing,hudson1999atomic}, and quasiparticle interface~\cite{hoffman2002imaging,mcelroy2003relating}.
In addition, STS has been extensively applied to a wide range of unconventional superconductors in order to elucidate the gap symmetry, including cuprate superconductors~\cite{fischer2007scanning,maggio1995direct,hoffman2002imaging,mcelroy2003relating,hanaguri2007quasiparticle,renner1998observation,wei1998directional,kashiwaya2000tunnelling,pan2001microscopic,lang2002imaging,schmidt2011electronic}, iron-based superconductors~\cite{hoffman2011spectroscopic,yin2009scanning,hanaguri2010unconventional,liu2018robust,machida2019zero}, graphene-based systems~\cite{choi2019electronic,oh2021evidence}, and heavy-fermion superconductors such as UTe$_2$~\cite{jiao2020chiral,aishwarya2023magnetic,gu2023detection,gu2025pair,wang2025imaging,wang2024observation,sharma2025observation,yang2025spectroscopic}.
In particular, the observation of in-gap states has provided compelling signatures of surface Andreev bound states (ABSs) that are intimately tied to topological superconductivity~\cite{hu1994midgap,sato2009topological,sato2010topological,sato2011topology,tanaka12,alicea2012new,miz16,sato2016majorana}.

A more recent development in STM is the use of $s$-wave superconducting tips~\cite{pan1998vacuum}.
One key advantage of using a superconducting tip lies in the singularities of its DOS, i.e., coherence peaks, which provide exceptionally high energy resolution in STS.
Furthermore, beyond single-particle spectroscopy, when both the tip and the sample are $s$-wave superconducting, STM enables the direct visualization of the condensate flow, observed as a Josephson current, giving rise to the so-called Josephson STM~\cite{naaman2001fluctuation,vsmakov2001josephson,rodrigo2004use,kimura2009josephson,ast2016sensing,randeria2016scanning,jack2016critical,ko2021statistical}.
Using Josephson STM, spatially nonuniform superconductivity---namely pair-density waves---has been successfully observed in cuprate-based superconductor~\cite{hamidian2016detection}, iron-based superconductors~\cite{cho2019strongly}, transition-metal dichalcogenides~\cite{liu2021discovery}, and kagome systems~\cite{deng2024chiral}.

While superconducting tips have been widely applied to $s$-wave superconducting samples, their potential for uncovering the nature of topological superconductors remains largely unexplored. 
Moreover, most existing theoretical studies of STM and STS with superconducting tips have primarily focused on the Josephson current, whereas the tunneling current originating from topological surface states has received little attention~\cite{graham2017imaging,graham2019josephson,choubey2024theory}.
This motivates the present study, in which we aim to clarify the elementary tunneling characteristics of STS with superconducting tips probing topological superconductors, focusing on both the differential conductance ($dI/dV$) and current noise.

The problem of electron tunneling in superconductor-superconductor (S-S) junctions has been extensively studied since the early days, particularly in the case where both electrodes are conventional superconductors~\cite{josephson1962possible,Andreev1964,ambegaoker1963tunneling,Werthamer1966nonlinear,klapwijk1982explanation,octavio1983subharmonic,ambegaokar1982quantum,eckern1984quantum,bratus1995theory,averin1995ac,cuevas1996hamiltonian,golubov2004current,vallet2024anderson,lahiri2025origin}.
At low bias voltage, below the superconducting gap, quasiparticle tunneling is absent, and two processes remain relevant.
The first is the Josephson effect, where Cooper pairs tunnel through the junction; when coupled to the electromagnetic environment, this process can become inelastic and manifests as a zero-bias conductance peak~\cite{anchenko1969josephson,ambegaoker1963tunneling,ingold1994cooper,harada1996cooper}.
The second is multiple Andreev reflection (MAR), in which an electron incident from one side can traverse the junction by undergoing successive Andreev reflections at the interface~\cite{klapwijk1982explanation,octavio1983subharmonic,bratus1995theory,cuevas1996hamiltonian}.
This MAR process gives rise to a characteristic subgap peak structure in the $dI/dV$, which has been observed experimentally~\cite{ternes2006subgap}.

We now turn to the case in which one of the electrodes is a topological superconductor.
In this situation, single-particle tunneling is likewise suppressed.
Moreover, since most topological superconductors are spin-triplet in nature, the Josephson tunneling current is expected to be negligible when the spin–orbit coupling is weak, owing to the orthogonality of the spin states between spin-singlet and spin-triplet superconductors~\cite{pal77,fenton,larkin,millis1988quasiclassical}.
The essential difference from conventional S-S junctions is that a topological superconductor hosts quasiparticle states 
on surfaces at arbitrary energies.
As a consequence, Andreev reflection can occur at any bias voltage, in which electrons and holes in the topological superconductor are converted into Cooper pairs in the $s$-wave superconductor and vice versa.
In particular, this process becomes dominant in the bias regime below the $s$-wave superconducting gap amplitude, and Andreev spectroscopy is accessible in this low-bias regime.

To elucidate Andreev spectroscopy of topological superconductors with superconducting tips, in this paper, we theoretically investigate a S-S junction between an $s$-wave superconductor and a topological superconductor.
We develop a general framework to describe tunneling between these two systems by employing the real-time Keldysh path-integral formalism, following Refs.~\cite{eckern1984quantum,schon1990quantum}.
We employ two approaches: the first one is an effective action method applicable to the weak tunneling regime, and the second one is a full-order Green function method applicable to general cases, including both the weak tunneling and strong tunneling regimes.
For the former approach,
we expand the tunneling Hamiltonian up to fourth order, which allows us to derive an effective action for a superconducting quantum circuit that incorporates Andreev reflection processes.
By analyzing the resulting effective action within the appropriate saddle-point approximation, we obtain an analytical expression for the tunneling current.
On the other hand, from the well-established case of normal metal-superconductor (N-S) junctions, it is known that MAR at the interface can lead to qualitatively different current-voltage characteristics in the strong-tunneling regime~\cite{blonder1982transition,daghero2010probing,kashiwaya2011edge}.
Motivated by this, we use the second approach, performing numerical calculations of the tunneling current in which all tunneling processes are incorporated as a self-energy.
This approach allows us to capture the current–voltage characteristics across the entire range of tunneling strengths, from the weak and intermediate regimes relevant to STM measurements to the strong-tunneling regime applicable to quantum point-contact experiments.
We note, however, that experimental access to the strong-tunneling regime is often challenging, since strong tunneling tends to cause charge accumulation at the tip apex, which can suppress superconductivity~\cite{naaman2001fluctuation}.

In addition to the tunneling current, we derive an analytical expression for the current noise and also calculate the current noise numerically.
At low temperatures, the dominant contribution to the noise arises from shot noise, which originates from the discrete and stochastic nature of the tunneling process and provides information about the effective charge involved in transport.
In particular, in the low-bias regime below the superconducting gap of the $s$-wave tip, the shot noise yields a Fano factor of $F = 2$, reflecting the $2e$ charge transfer characteristic of Andreev reflections~\cite{blanter2000shot,kobayashi2021shot,beenakker1997random,kokkeler2025full}.

We systematically consider various types of topological superconductors, as summarized in Table~\ref{list_SC}.
The essential feature of topological superconductivity lies in the odd-parity nature of the gap function, which tightly links the pairing symmetry to the surface orientations where ABSs emerge. Consequently, the observation of such surface states by Andreev spectroscopy can provide direct insights into the symmetry of the superconducting order parameter.
To clarify this connection, we perform comprehensive calculations of the surface DOS, the $dI/dV$ spectra, and the current noise for symmetric surface such as $(100)$, $(010)$, and $(001)$ surface, using the standard models of topological superconductors. 
To simulate realistic experimental conditions, we include the effects of finite temperature and quasiparticle lifetime broadening (Dynes broadening) by appropriately setting the parameters~\cite{dynes1978direct}.
In particular, the residual DOS induced by Dynes broadening gives rise to single-electron tunneling even within the superconducting gap, leading to a competition between ${|T|}^2$ and ${|T|}^4$ tunneling processes~\cite{niu2024why}.
Here, $|T|$ is a tunneling amplitude.
This competition can be directly revealed by calculating the Fano factor, which highlights the crucial role of the $s$-wave tip quality in Andreev spectroscopy.
Our results will provide useful guidance for future STS experiments with superconducting tips through Andreev spectroscopy and noise spectroscopy, offering a potential route to identify the pairing symmetry of topological superconductors.

The rest of this paper is organized as follows.
In Sec.~\ref{Sec2}, we summarize the theoretical formulation.
Section~\ref{2A} presents the derivation of the effective tunneling action for a junction between an $s$-wave superconductor and a topological superconductor in the weak-tunneling limit, based on the real-time Keldysh path-integral formalism.
This derivation follows the approach of Refs.~\cite{eckern1984quantum,schon1990quantum}, extended to incorporate Andreev reflection processes, with the full details provided in Appendix~\ref{Appendix1}.
In Sec.~\ref{2D}, we give a brief introduction to noise measurements and derive analytical expressions for the current noise from the tunneling action.
In Sec.~\ref{2B}, we describe the numerical algorithm used to compute the tunneling current and current noise, including perturbations to all orders.
In Sec.~\ref{Sec3}, we summarize the representative topological superconductors, including their topological surface states.
Section~\ref{Sec4} presents the numerical results for the $dI/dV$ spectra and noise characteristics, simulating STS measurements with an $s$-wave superconducting tip. 
Section~\ref{Sec6} is devoted to the conclusion and discussion.

\section{Analytical Expressions for the Tunneling Current and Noise}  \label{Sec2}
In this section, we develop analytical frameworks to describe the tunneling current and current noise in a voltage-biased S-S junction, starting from a microscopic Hamiltonian.
We employ the Keldysh path integral formalism to describe the nonequilibrium tunneling processes.
In Sec.~{\ref{2A}}, we construct an effective tunneling action by expanding the tunneling Hamiltonian up to fourth order, which enables us to incorporate Andreev reflection processes in addition to single-particle tunneling and Cooper-pair tunneling (Josephson) processes.
Applying the saddle-point approximation to the resulting action, we obtain an analytic Kirchhoff's equation that governs the dynamics of the superconducting circuit.
In Sec.~\ref{2D}, we analyze the current-noise characteristics of the S-S junction.

\subsection{Tunneling action and current\label{2A}}
\begin{figure}[t]
    \centering
    \includegraphics[width=\linewidth]{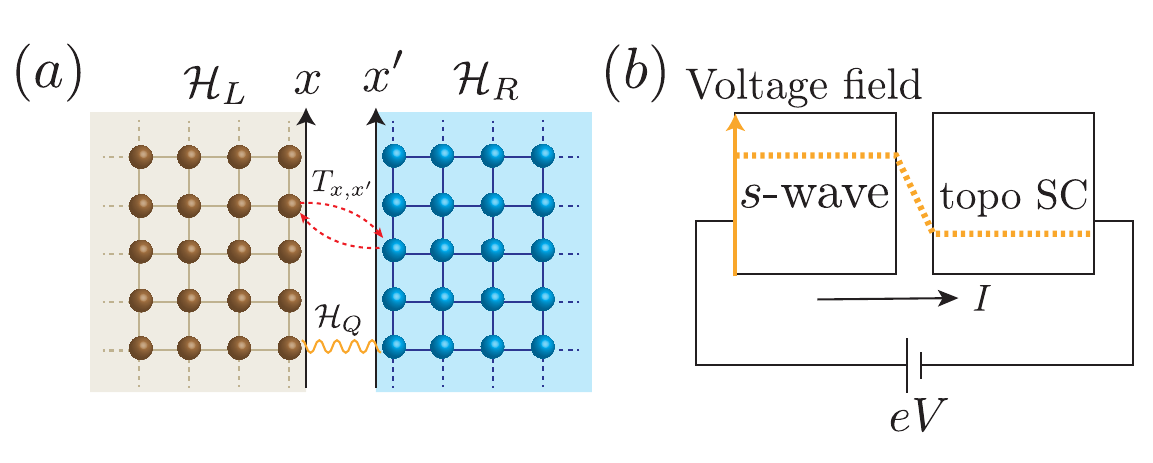} 
    \caption{(a)~Schematic illustration of the full Hamiltonian, including electron tunneling between the left and right lattices.
    $x$ denotes the coordinate along the interface, and for spatially local tunneling, as in STM, the tunneling matrix element is taken as $T_{x,x'} = T \delta(x)\delta(x')$.
    (b)~Circuit diagram of a superconducting junction.}
    \label{fig:setup}
\end{figure}

Let us consider two superconducting systems that are weakly coupled and connected to an external voltage source, as shown in Fig.~\ref{fig:setup}.
Throughout this paper, we assume that the left electrode, corresponding to the STM tip, is an $s$-wave superconductor, whereas the right electrode, representing the superconducting sample, can be either an $s$-wave or an unconventional superconductor.
Such an S–S junction constitutes a macroscopic quantum system characterized by the phase difference between the two superconductors.
The goal of this section is to derive an effective action that governs the dynamics of phase difference, starting from the microscopic Hamiltonian.

We begin with the total microscopic Hamiltonian
\begin{eqnarray}
    \label{eq:total}
    \mathcal{H}_{\rm tot} = \mathcal{H}_L + \mathcal{H}_R + \mathcal{H}_T +\mathcal{H}_Q,
\end{eqnarray}
where
\begin{align}
    \label{eq:HL}
    &\mathcal{H}_L =  \int d{\bm r}~\psi_{L\sigma}^{\dagger}({\bm r})\Big[-\frac{\hbar^2}{2m}\bm{\nabla}^2-\mu\Big]\psi_{L\sigma}({\bm r}) + \mathcal{H}_{\rm pair},  \\
    &\mathcal{H}_R = \mathcal{H}_L(L\leftrightarrow R), \\
    &\mathcal{H}_Q = \frac{e^2}{8C}(N_L-N_R)^2, \\
    \label{eq:HT}
    &\mathcal{H}_T = \int_{{\bm r}\in L}d{\bm r}\int_{{\bm r}'\in R}d{\bm r}'~\psi_{L\sigma}^{\dagger}({\bm r})T_{{\bm r},{\bm r}'}\psi_{R\sigma}({\bm r}') + \rm{h.c.},
\end{align}
where $\psi^{\dag}_{L(R)\sigma}({\bm r})$ and $\psi_{L(R)\sigma}({\bm r})$ are the creation and annihilation operators of electrons in the left (right) electrodes. The BCS Hamiltonian for the left (right) electrodes, $\mathcal{H}_{L(R)}$, consists of the noninteracting part and the pairing interaction $\mathcal{H}_{\rm pair}$, where $m$ is the mass of electrons and $\mu$ is the chemical potential.
$\mathcal{H}_{Q}$ describes the effective capacitive interaction at the junction, which originates from the Coulomb interaction between the left and right superconducting electrodes. Here, $C$ is the capacitance depending on the geometry and properties of the junction, and $N_{L(R)}\equiv \sum_{\sigma} \int d{\bm r}\psi^{\dag}_{L(R)\sigma}({\bm r})\psi_{L(R)\sigma}({\bm r})$ is the electron number of the left (right) side.
The tunneling Hamiltonian, $\mathcal{H}_{T}$, describes electron transfer between the two superconductors.
The tunneling matrix $T_{r,r'}$ reflects the geometry of the junction.
In systems such as STM, where electron tunneling occurs locally in space, the uncertainty principle between position and momentum implies that the tunneling matrix can be regarded as momentum independent, as $T$.

To treat the nonequilibrium dynamics of the S-S junction, we employ the real-time path integral formalism~\cite{eckern1984quantum,schon1990quantum}.
Following the Keldysh contour approach, the partition function is expressed as
\begin{align}
    Z_K = \mathrm{Tr}_{\psi,\bar{\psi}}\exp[\frac{i}{\hbar}S_{\rm tot}],
\end{align}
where $S_{\rm tot}$ is the total action expressed in terms of the Grassmann fields $\psi$ and $\bar{\psi}$, and is defined as a functional integral along the Keldysh contour $\mathcal{C}$.
The trace $\mathrm{Tr}_{\psi,\bar{\psi}}$ denotes the functional integral over these fields on the contour.

Here, we briefly outline the procedure for obtaining effective action, with the full details provided in Appendix~\ref{Appendix1}. The derivation process involves the following four steps.
(1)~We first introduce collective variables via Hubbard-Stratonovich transformation of the interacting terms:
the complex superconducting order-parameter fields $\Delta_L$ and $\Delta_R$, and a real voltage field $V$.
(2)~By integrating out the fermionic degrees of freedom, we obtain an action expressed solely in terms of the collective variables.
(3)~As a zeroth-order approximation in the tunneling Hamiltonian, we assume that the left and right superconductors are in equilibrium and perform a saddle-point analysis.
The saddle-point equation for the amplitude of the order-parameter field reproduces the gap equation,
while that for the voltage field yields the Josephson relation,
\begin{align}
    \label{eq:Josephson}
    \dv{\phi}{t} = \frac{2e}{\hbar}V,
\end{align}
where $\phi$ is the superconducting phase difference.
By neglecting amplitude and voltage fluctuations around these mean-field solutions, the only remaining low-energy collective variable is the superconducting phase difference.
(4)~Finally, by perturbatively expanding the action in the tunneling matrix---which couples to the phase difference through a gauge transformation---we obtain an effective action describing the phase dynamics.
This perturbative expansion allows us to classify the contributions from different tunneling processes: single-particle tunneling, Josephson tunneling, and Andreev reflection processes.

As a result, we obtain the effective tunneling action:
\begin{widetext}
    \begin{align}
        \label{eq:Action}
        S_{\rm tun}[\phi,\chi] =& -\frac{\hbar}{2e}\int_{-\infty}^{\infty} dt~  \bigg[\frac{C}{2e}\pdv[2]{\phi}{t} + I_{\rm c}\sin\phi - {I_{\rm ex}} \bigg]\chi \nn \\
        &- \frac{\hbar}{2e}\int_{-\infty}^{\infty} dtdt'~\bigg[ 2e\alpha^I(t-t')\sin(\frac{\phi-\phi'}{2})\chi - \frac{1}{2}e\alpha^k(t-t')\cos(\frac{\phi-\phi'}{2})\chi\chi' \bigg] \nn \\
        &-\frac{\hbar}{2e}\int_{-\infty}^{\infty} dtdt'~\bigg[2e\gamma^I(t-t')\sin(\phi-\phi')\chi - e\gamma^k(t-t')\cos(\phi-\phi')\chi\chi'\bigg],
\end{align}
\end{widetext}
where we employ the classical-quantum representation:
$\phi$ and $\chi$ denote the classical and quantum components, respectively,
defined as $\phi = (\phi^+ + \phi^-)/2$ and $\chi = \phi^+-\phi^-$,
with $\phi^{+(-)}$ denoting the fields on the forward (backward) branch of the Keldysh contour. 
For simplicity in notation, we use the shorthand $\phi=\phi(t)$, $\phi'=\phi(t')$, $\chi=\chi(t)$, and $\chi'=\chi(t')$.
The quantities $I_{\rm c}$, $\alpha^{I(K)}$, and $\gamma^{I(K)}$ correspond to tunneling self-energies obtained from the Green’s functions, as discussed below.

The first line of Eq.~\eqref{eq:Action} contains the capacitive term,
the Josephson coupling term characterized by the critical Josephson current $I_{\rm c}$,
and the contribution from the external current $I_{\rm ex}$.
The critical current $I_c$ is obtained from the frequency summation of the product of the anomalous Green's function of the left and right superconductors~(see Eqs.~\eqref{eq:beta_r}, \eqref{eq:beta_r}, and \eqref{eq:Ic}).
It should be noted that in the case of the S-S junctions between an anisotropic superconductor and an $s$-wave superconductor, the Josephson coupling is generally weak due to the symmetry mismatch.
For this reason, Josephson currents lie outside the main scope of the present study and do not appear in the subsequent calculations.
However, spin–orbit coupling and the breaking of local translational symmetry can give rise to a finite Josephson coupling~\cite{fenton1986josephson,millis1988quasiclassical,hasegawa1998josephson,choubey2024theory}.

The terms described by the kernels $\alpha^{I(K)}$ and $\gamma^{I(K)}$ correspond to the contribution from single-particle tunneling and Andreev reflection, respectively.
For each contribution, the terms linear in $\chi$ in Eq.~\eqref{eq:Action} represent dissipation arising from incoherent tunneling processes, i.e., the tunneling currents, 
whereas the quadratic terms in $\chi$ account for the current noise.
Since the current noise will be discussed in detail in Sec.~\ref{2D}, we neglect the quadratic terms at this stage and focus on the contributions linear in $\chi$.
The kernel $\alpha^I$ describes the phase relaxation induced by single-particle tunneling processes and can be expressed in terms of the tunneling current as
\begin{align}
    \alpha^I(\omega) =& \frac{i}{2e}I_{\rm qp}(\hbar\omega/e), 
\end{align}
The explicit expressions for the tunneling currents are given by the well-known expression
\begin{align}
    I_{\rm qp}(V) = (4\pi |T|^2 e/\hbar)\int_{-\infty}^{\infty}dE~\rho_L(E-eV)\rho_R(E)\nn \\
    \label{eq:Isingle}
    \times\{n_{\rm F}(E-eV) - n_{\rm F}(E) \}.
\end{align}
Here, $\rho_{L(R)}$ denotes quasiparticle DOS in the left (right) electrode,
and $n_{\rm F}(x) = 1/(e^{x/k_{\rm B}\Theta}+1)$ is the Fermi distribution function,
with $k_{\rm B}$ the Boltzmann constant and $\Theta$ the temperature.
Similarly, the kernel $\gamma^I$ characterizes phase relaxation due to Andreev reflection, and is related to the Andreev current via
\begin{align}
    \gamma^I(\omega) =& \frac{i}{2e}I_{\rm AR}(\hbar\omega/2e).
\end{align}
The explicit expression for the Andreev current is given by
\begin{widetext}
\begin{align}
    I_{\rm AR}(V) =&  (4\pi^3|T|^4eN_L^2/\hbar)\int_{-\infty}^{\infty}dE~\rho_R(-E+eV)\rho_R(E+eV)\nn\\
    &\hspace{100pt}\times\Big[~|F_L(E)|\{n_{\rm F}(E-eV)-n_{\rm F}(E) \} +F_L(E)\{n_{\rm F}(E)-n_{\rm F}(E+eV) \}~\Big],
    \label{eq:IAndreev}
\end{align}
\end{widetext}
where $N_{L(R)}$ is the normal state DOS at the Fermi level,
and $F_L(E) = \Delta_L^2/(\Delta_L^2 - E^2)$ is the pair amplitude of the left $s$-wave superconductor.
This expression reflects the fact that a Cooper pair in the STM tip (left electrode) tunnels into the sample superconductor (right electrode) as an electron-hole pair, or vice versa.

We remark that when the right electrode is a topological superconductor and hosts in-gap states, Eqs.~\eqref{eq:Isingle} and \eqref{eq:IAndreev} already provide a reliable description of the tunneling processes over the entire bias range, as we explain below.
For the bias voltage $eV>\Delta_L$, the single particle current in Eq.~\eqref{eq:Isingle} gives the leading contribution, as it represents a second-order $(|T|^2)$ tunneling process.
In contrast, for $eV<\Delta_L$, this contribution becomes strongly suppressed because the $s$-wave electrode possesses a fully gapped quasiparticle DOS.
In this low-bias regime, the next-to-leading process described by Eq.~\eqref{eq:IAndreev}---a fourth-order $(|T|^4)$ tunneling contribution---remains finite for any nonzero bias due to the presence of in-gap states in the topological superconductor.
Although a complementary process, obtained by exchanging $L$ and $R$ in Eq.~\eqref{eq:IAndreev}, also exists,
its contribution is proportional to $\rho_{L}\rho_{L}$ and is therefore subject to strong suppression from the fully gapped DOS of the $s$-wave electrode. 
By contrast, when the right electrode is also a fully gapped superconductor, such as an $s$-wave superconductor, Eq.~\eqref{eq:IAndreev} alone is not sufficient.
Because Eq.~\eqref{eq:IAndreev} also becomes suppressed by the fully gapped quasiparticle DOS, the complementary Andreev process obtained by exchanging $L$ and $R$ cannot be neglected, as it contributes at the same order.
Moreover, in the low-bias regime, even these $|T|^4$ processes become strongly suppressed, and higher-order MAR processes dominate the subgap transport (see Sec.~\ref{5B}).

That $\alpha^I$ and $\gamma^I$ represent tunneling currents can be demonstrated explicitly by performing a saddle-point approximation with respect to $\chi$ in Eq.~\eqref{eq:Action}.
This procedure yields the semiclassical equation of motion for the phase difference,
\begin{eqnarray}
    \label{eq:RCSJ}
    \frac{C\hbar}{2e}\pdv[2]{\phi}{t} + I_{\rm qp}(V) + I_{\rm AR}(V) + I_{\rm c}\sin\phi = I_{\rm ex},
\end{eqnarray}
where we have used the Josephson relation Eq.~\eqref{eq:Josephson}.
This equation corresponds to Kirchhoff's current law for the resistively and capacitively shunted junction (RCSJ) model, extended here to include Andreev reflection current $I_{\rm AR}$.

\subsection{Noise measurements\label{2D}}
While tunneling current provides information on the time-averaged electron transport, noise measurements offer a more detailed physical picture of the nonequilibrium transport processes.
In this subsection, we give a brief introduction to current noise and summarize noise properties probed by superconducting-tip STM~\cite{blanter2000shot,kobayashi2021shot,beenakker1997random,kokkeler2025full}.
Noise can be classified into two main categories according to its origin: thermal noise and shot noise.
Thermal noise arises from thermal excitations at finite temperature, whereas shot noise originates from the discrete and probabilistic nature of tunneling processes.
In this work, we focus primarily on shot noise, which is briefly introduced below.

We consider elementary tunneling events, in which a particle tunnels with probability $\mathcal{T}$ and is reflected with probability $1-\mathcal{T}$, each transferring an effective charge $e^*$.
For instance, in single particle tunneling, we have $e^*=e$, whereas in Andreev reflection processes, $e^*=2e$.
The time-averaged current is then proportional to the tunneling probability, $\langle I(t)\rangle\propto e^*\mathcal{T}$.
Assuming that the junction is weak and tunneling events are rare, the statistics of tunneling follow a Poisson process.
In a Poisson process, the variance of the number of tunneling events equals its mean, which leads to the noise power $P_N\propto 2{e^*}^2\mathcal{T}$.
Here, the noise power is defined as $P_N=\langle\Delta I(t)^2 \rangle$, where $\Delta I(t)\equiv I(t)-\langle I(t)\rangle$ is the fluctuation of the current from its time-averaged value.
Introducing the Fano factor as the ratio of noise power to current,
\begin{eqnarray}
     F\equiv \frac{P_N}{2e|I|} = e^*/e.
\end{eqnarray}
We find that noise measurements provide direct access to the effective charge involved in the tunneling processes.
In superconducting junctions, noise measurements revealing an effective charge of $2e$ have been reported in several systems, including mesoscopic junctions~\cite{jehl2000detection,kozhevnikov2000observation,chen2012excess}, quantum wires~\cite{ronen2016charge}, and even superconducting-tip STM~\cite{bastiaans2019imaging,ge2023single,ge2024direct}.

The microscopic derivation of noise has been largely completed in Sec.~\ref{2A}.
In the Keldysh action, the quadratic terms of the quantum component, i.e., the $\chi\chi^{\prime}$ terms in Eq.~\eqref{eq:Action}, correspond to the contribution of noise.
It can be decoupled via the Hubbard-Stratonovich transformation,
which introduces a Langevin noise source $I_{N}(t)=I_{N}^{\rm qp}(t)+I_{N}^{\rm AR}(t)$ in the Kirchhoff current equation of Eq.~\eqref{eq:RCSJ}.
The corresponding noise action is 
\begin{eqnarray}
    S_{\rm Noise} = -\frac{\hbar}{2e}\int_{-\infty}^{\infty} dt~\Big[I_N^{\rm qp}(t)+I_N^{\rm AR}(t)\Big]\chi ,
\end{eqnarray}
where $I_N^{\rm qp}$ and $I_N^{\rm AR}$ are constructed as appropriate Gaussian random variables, whose variance yields
\begin{gather}
    \label{eq:noise_qp}
    \langle I_N^{\rm qp}(t)I_N^{\rm qp}(t')\rangle = -2ie^2\alpha^k(t-t')\cos\frac{\phi-\phi'}{2}, \\
    \label{eq:noise_AR}
    \langle I_N^{\rm AR}(t)I_N^{\rm AR}(t')\rangle = -4ie^2\gamma^k(t-t')\cos({\phi-\phi'}).
\end{gather}
Importantly, we note that there is no cross-correlation between the single-particle  and Andreev reflection contributions.
We define noise power as
\begin{eqnarray}
    P_N(\omega) = 2\int_{-\infty}^{\infty} dt~e^{i\omega t}\langle I_N(t)I_N(0)\rangle.
\end{eqnarray}
Using Eq.~\eqref{eq:noise_qp}, the DC noise power arising from single particle tunneling process is given by
\begin{eqnarray}
    \label{eq:PNqp}
    P_N^{\rm qp} = 2e^2(-i)\alpha^k(eV/\hbar).
\end{eqnarray}
Similarly, the contribution from the Andreev reflection, based on Eq.~\eqref{eq:noise_AR}, is expressed as
\begin{eqnarray}
    \label{eq:PNAR}
    P_N^{\rm AR} = 2(2e)^2(-i)\gamma^k(2eV/\hbar).
\end{eqnarray}
The total noise is $P_N = P_N^{\rm qp} + P_N^{\rm AR}$.
The noise kernels $\alpha^k$ and $\gamma^k$ are related to the corresponding current $\alpha^I$ and $\gamma^I$ through the fluctuation-dissipation relation,
\begin{gather}
    \alpha^k(eV/\hbar) = \alpha^I(eV/\hbar)\coth(eV/k_{\rm B}\Theta), \\
    \gamma^k(e^*V/\hbar) = \gamma^I(e^*V/\hbar)\coth(e^*V/k_{\rm B}\Theta),
\end{gather}
where $e^*=2e$ for $\gamma$.

First, let us consider the limit $eV\gg k_{\rm B}\Theta$ (shot-noise regime).
In this case, $\coth(eV/k_{\rm B}\Theta)\rightarrow 1$, so that the Keldysh components of the kernels $\alpha$ and $\gamma$ coincide with their dissipative counterparts.
This implies that the time-averaged current and the noise power are governed by the same kernels, and hence the Fano factor directly reflects the effective charge involved in the tunneling process.
In realistic situations where both single particle tunnel and Andreev reflection processes coexist, the total Fano factor lies between $1$ and $2$, $1\le F \le 2$, depending on the relative contributions of each tunneling channel.
On the other hand, in the limit $eV\ll k_{\rm B}\Theta$ (thermal-noise regime), the noise is enhanced as $\coth(eV/k_{\rm B}\Theta)\approx k_{\rm B}\Theta/eV$.
In particular, since thermal noise remains finite even at zero bias, the Fano factor at finite temperature diverges as $V\rightarrow 0$.

\section{
Numerical Method for Nonperturbative Calculation of Tunneling Current and Noise
\label{2B}}
In the previous section, we analytically derived the expressions for the tunneling action arising from single-particle and Andreev reflection processes.
Beyond the weak-tunneling regime, however, one must take into account multiple reflections at the interface to obtain the tunneling current.
Cuevas {\it et al.} employed the Keldysh Green's function formalism to 
obtain the full tunneling current and noise for arbitrary transmission and bias in $s$-wave/$s$-wave junctions by incorporating MAR through a renormalized tunneling matrix~\cite{cuevas1996hamiltonian,cuevas1999shot}.
In this system, the interface is governed solely by Andreev reflection, which makes an analytic treatment possible.
In contrast, when one of the electrodes is a topological superconductor, normal particle reflection is also allowed in addition to Andreev reflection. 
This considerably complicates the analysis, and numerical calculations become indispensable. 
In the following, we describe the numerical procedure employed, while further technical details are provided in Ref.~\cite{tei2025topological}.

Based on the microscopic Hamiltonian given in Eqs.~\eqref{eq:HL}-\eqref{eq:HT}, the current operator is expressed as
\begin{gather}
    I = \frac{ie}{\hbar} \sum_{\bm{k},\bm{k'}}~\Big\{c_{L\bm{k}}^{\dagger}Tc_{R\bm{k'}} - c_{R\bm{k'}}^{\dagger}T^\dagger c_{L\bm{k}}\Big\}.
\label{eq:Iop}
\end{gather}
Here, $c^{\dagger}_{L\bm{k}}~(c_{L\bm{k}})$ and $c^{\dagger}_{R\bm{k}}~(c_{R\bm{k}})$ denote the creation (annihilation) operators for electrons in the left and right electrodes, respectively, with momentum $\bm{k}$ parallel to the surface.
We further assume local tunneling, such that the tunneling matrix $T$ is independent of momentum.

Under a finite bias voltage, the expectation value of the tunneling current in Eq.~\eqref{eq:Iop} can be evaluated within the Keldysh Green's function formalism as
\begin{eqnarray}
    \label{eq:Ifull}
    \langle I(t)\rangle = -e~\mathrm{Tr} [\sigma_zT  G_{LR}^<(t,t) - \sigma_zT^\dagger G_{RL}^<(t,t)],
\end{eqnarray}
where $\sigma_z$ denotes tha Pauli $z$ matrix in Nambu space.
Here, $G_{LR}^<$ and $G_{RL}^<$ denote the lesser components of the junction Green's functions,
which describe particle propagation from the left to the right system and vice versa.
In Nambu space, they are defined as
\begin{gather}
    G^<_{ij}(t,t') \equiv \sum_{\bm{k}, \bm{k}'} G^<_{ij}({\bm k},t;{\bm k}',t'), \nn \\ 
    G^<_{ij}({\bm k},t;{\bm k}',t') = \frac{i}{\hbar}\begin{pmatrix}
         \langle \tilde{c}_{j\bm{k}'}^{\dagger}(t')\tilde{c}_{i\bm{k}}(t)\rangle & \langle \tilde{c}_{j\bm{k}'}(t')\tilde{c}_{i\bm{k}}(t)\rangle \\
         \langle \tilde{c}_{j\bm{k}'}^{\dagger}(t')\tilde{c}_{i\bm{k}}^{\dagger}(t)\rangle & \langle \tilde{c}_{j\bm{k}'}(t')\tilde{c}_{i\bm{k}}^{\dagger}(t)\rangle
    \end{pmatrix},
\end{gather}
with $i,j = L,R$.
The operators $\tilde{c}$ and $\tilde{c}^\dagger$ incorporate the effect of the bias voltage, treated as a chemical potential difference ($eV = \mu_L-\mu_R$), via a gauge transformation:
\begin{align}
    \label{eq:cL_eV}
    \tilde{c}_L(t) &= c_L(t) e^{-i e V t / \hbar}, & \tilde{c}_L^\dagger(t) &= c_L^\dagger(t) e^{i e V t / \hbar}, \\
    \tilde{c}_R(t) &= c_R(t), & \tilde{c}_R^\dagger(t) &= c_R^\dagger(t).
\end{align}
The tunneling matrix in Nambu space is given by 
\begin{eqnarray}
    T = \begin{pmatrix}
        |T|e^{i\phi_0} & \\
        & -|T|e^{-i\phi_0}
    \end{pmatrix},
\end{eqnarray}
where $\phi_0$ denotes the static phase difference between the two electrodes.
We now focus on the dc components of the tunneling current, defined as $I \equiv \overline{I(t)}$.

In addition, the (symmetrized) current noise power is defined as
\begin{eqnarray}
    \label{eq:Noisefull}
        P_N(\omega,t) &=& \int_{-\infty}^{\infty} dt'  e^{i\omega t'}\langle \{ \delta I(t+t'),\delta I(t) \} \rangle \nn \\
    &\equiv& \int_{-\infty}^{\infty} dt' e^{i\omega t'}\chi (t,t+t'), 
\end{eqnarray}
where $\delta I(t) = I(t) - \langle I(t)\rangle $ denotes the current fluctuation.
Here, we have introduced the current--current correlation function as $\chi(t,t') \equiv \langle \{I(t),I(t')\}\rangle$.
Using the Keldysh Green's functions, the correlation function can be expressed as 
\begin{align}
    \chi(t, t') = 2e^2 \mathrm{Tr} \Big\{ 
    &-T\sigma_zG_{LR}^>(t,t')T\sigma_zG_{LR}^<(t',t) \nn \\
    & -T^\dagger\sigma_z G_{RL}^>(t,t')T^\dagger\sigma_z G_{RL}^<(t',t)  \nn \\
    &+ T\sigma_zG_{LL}^>(t,t')T^{\dagger}\sigma_zG_{RR}^<(t',t) \nn \\
    & +T^\dagger\sigma_z G_{RR}^>(t,t')T\sigma_zG_{LL}^<(t',t) \Big\} \nn \\
    &+ (t\leftrightarrow t'). \label{eq:IIcorr}
\end{align}
Here, $G_{ij}^>$ denotes the greater components of the junction Green's function, which has the relation
$G_{ij}^>(t,t') = [G_{ji}^<(t',t)]^\dagger$.
We now focus on the the dc components of the tunneling current and the current noise. 
The time-independent dc noise is defined as $P_N\equiv \overline{P_N(0,t)}$.

The remaining task is to evaluate the junction Green’s functions by solving the Dyson equation in Keldysh space.
Treating the tunneling matrix as a single-particle perturbation, we obtain the following set of Dyson equations:
\begin{widetext}
\begin{gather}
    \label{eq:GLL}
    \check{G}_{LL} = \check{G}_{LL}^{(0)} + \check{G}_{LL}^{(0)}\circ \check{T}^{\dagger} \circ \check{G}_{RL}=\check{G}_{LL}^{(0)} + \check{G}_{LR}\circ \check{T} \circ \check{G}_{LL}^{(0)}, \\
    \check{G}_{RR} = \check{G}_{RR}^{(0)} + \check{G}_{RR}^{(0)}\circ \check{T}^{\dagger} \circ \check{G}_{LR}=\check{G}_{RR}^{(0)} + \check{G}_{RL}\circ \check{T} \circ \check{G}_{RR}^{(0)}, \\
    \label{eq:GLR}
    \check{G}_{LR} = \check{G}_{LL}^{(0)} \circ \check{T} \circ \check{G}_{RL} = \check{G}_{LR}\circ \check{T} \circ \check{G}_{RR}^{(0)}, \\
     \label{eq:GRL}
    \check{G}_{RL} = \check{G}_{RR}^{(0)} \circ \check{T} \circ \check{G}_{LR} = \check{G}_{RL}\circ \check{T} \circ \check{G}_{LL}^{(0)}.
\end{gather}
\end{widetext}
Here, $\check{G}^{(0)}_{LL}$ and $\check{G}^{(0)}_{RR}$ denote the non-perturbative Green's function for the isolated superconducting systems, corresponding to Hamiltonian $\mathcal{H}_L$ and $\mathcal{H}_R$, respectively.
The ``check" symbol $(\check{M})$ indicates that the matrix $M$ is expressed in the rotated Keldysh basis,
and the circle product $\circ$ denotes integration over the internal variables, such as time and momentum.
By substituting Eq.~\eqref{eq:GRL} into Eq.~\eqref{eq:GLR}, we obtain
\begin{gather}
    \label{eq:Dyson_GLR}
    \check{G}_{LR} = \check{G}_{LL}^{(0)}\circ\check{\Sigma}\circ\check{G}_{RR}^{(0)}, \\
    \label{eq:Dyson_sigma}
    \check{\Sigma} \equiv \check{T}^{\dagger} + \check{T}^{\dagger}\circ\check{G}_{RR}^{(0)}\circ \check{T}\circ \check{G}_{LL}^{(0)} \circ\check{\Sigma}.
\end{gather}
The self-energy $\check{\Sigma}$ introduced here accounts for multiple reflections at the interface.
By solving the Dyson equation for the self-energy in Fourier space, one obtains the full Green's function (see Appendix~\ref{Appendix2}).
This procedure allows us to evaluate the tunneling current in Eq.~\eqref{eq:cL_eV} and the current noise in Eq.~\eqref{eq:Noisefull}, including contributions from tunneling processes to all orders.

To facilitate the discussion below, we introduce a dimensionless transparency as a measure of the tunneling strength.
From the Dyson equation, the tunneling current can be expressed, within the approximation that neglects the bias dependence, as a series in $\pi^2N_LN_R|T|^2$.
This motivates the definition of effective transparency,
\begin{eqnarray}
    \label{eq:alpha}
    \alpha = \frac{4\pi^2|T|^2N_LN_R}{(1+\pi^2|T|^2N_LN_R)^2}.
\end{eqnarray}
For conventional $s$-wave/$s$-wave junctions, this expression reproduces the exact transparency that includes all MAR processes~\cite{cuevas1996hamiltonian}.
Strictly speaking, in the present junction 
where one side is a topological superconductor, since both normal and Andreev reflections contribute, and the DOS exhibits strong low-energy structure,
Eq.~\eqref{eq:alpha} does not provide a strictly accurate description of the transparency in this case.
Nevertheless, since it still serves as a useful quantitative measure of the tunneling strength, we adopt this definition throughout this work.

\section{topological superconductors and surface states}
\label{Sec3}

\begin{table*}[t]
    \caption{Summary of topological superconducting states. ``Name'', ``Spin'', ``Gap function'', and ``Bulk node'' denote the name of the pairing states, their spin state, gap functions, and the nodal structures in bulk, respectively. ``Surface'' and ``ABSs'' are the surface orientation and the existence/absence of Andreev bound states, respectively, and the characteristic shape of the surface density of states is summarized in the column of ``sDOS''.}
    \begin{ruledtabular}
    \begin{tabular}{ccccccccc}
    &Name &Spin & Gap function & Bulk node & Surface & ABSs & sDOS & \\
    \hline
    &$s$ &singlet & $\psi = \Delta_0$ & Full gap &  & None & U-shaped &   \\ 
    \hline
    &BW $p$ & triplet & ${\bm d}=(\hat{k}_x,\hat{k}_y,\hat{k}_z)$ & Full gap & $(100)$,~$(010)$,~$(001)$ & Dirac cone & V-shaped  & \\ 
    \hline
    & Chiral $p$ &  triplet & ${\bm d}=(0,0,\hat{k}_x+i\hat{k}_y)$ & point node & \begin{tabular}{c} $(100)$,~$(010)$ \\ $(001)$ \end{tabular} 
    & \begin{tabular}{c} Fermi arc \\ None \end{tabular} 
    & \begin{tabular}{c} Flat-type \\ V-shaped \end{tabular} & \\
    \hline
    &Helical $p$ & triplet & ${\bm d}=(\hat{k}_x,\hat{k}_y,0)$ & point node 
    & \begin{tabular}{c} $(100)$,~$(010)$ \\ $(001)$ \end{tabular}
    & \begin{tabular}{c} Fermi arc \\ None \end{tabular}
    & \begin{tabular}{c} Flat-type \\ V-shaped \end{tabular} & \\
    \hline
    &Polar $p$ & triplet & ${\bm d}=(\hat{k}_z,0,0)$ & line node 
    & \begin{tabular}{c} $(001)$ \\  $(100)$,~$(010)$ \end{tabular} 
    & \begin{tabular}{c} Drum head \\ None \end{tabular} 
    & \begin{tabular}{c} Peak-type \\ V-shaped \end{tabular} & \\
    \hline
    & $d_{x^2-y^2}$ & singlet & $\psi = \hat{k}_x^2-\hat{k}_y^2$ & line node  
    & \begin{tabular}{c} $(100)$,~$(010)$,~$(001)$  \\ $(110)$ \end{tabular} 
    & \begin{tabular}{c} None \\   Drum head \end{tabular} 
    & \begin{tabular}{c} V-shaped  \\ Peak-type \end{tabular} &  \\
    \end{tabular}
    \end{ruledtabular}
    \label{list_SC}
\end{table*}

In this section, we summarize various types of topological superconductors.
The essential origin of topological properties in superconductors lies in their odd-parity pairing symmetry.
As a consequence, most spin-triplet superconductors are strong candidates for realizing topological superconductivity.
$d$-wave superconductors, on the other hand, possess an odd functional dependence of the gap along certain momentum directions, leading to the emergence of ABSs on specific crystal faces.
In this sense, most anisotropic superconductors can host ABSs.
In this paper, we focus on $p$-wave and $d$-wave superconductors.

The superconducting order parameter generally carries spin degrees of freedom and can be written in the following form
\begin{eqnarray}
    \hat{\Delta}(\bm{k}) = (\bm{d}(\bm{k})\cdot\boldsymbol{\sigma} + \psi(\bm{k}))i\sigma_y,
\end{eqnarray}
where $\psi(\bm{k})$ denotes the spin-single component and $\bm{d}(\bm{k})$ represents the spin-triplet component.
Due to the fermionic antisymmetry of the Cooper pair wave function, $\psi(\bm{k})$ must be an even function of momentum (e.g., $d$-wave), while $\bm{d}(\bm{k})$ must be odd (e.g., $p$-wave).

Table~\ref{list_SC} summarizes representative examples of topological superconducting states.
In this table, we include nodal superconductors such as a polar state and a $d$-wave pairing state in addition to fully-gapped topological superconductors, because these nodal superconducting states also support gapless ABSs protected by weak topological invariants~\cite{sato2011topology,yang2014dirac,sch15}.
Classifying the ABSs according to the surface DOS helps elucidate the characteristic features of tunneling spectroscopy.
The dimensionality and dispersion of ABSs have a significant impact on the resulting surface DOS.
Based on these properties, we classify the surface DOS into three characteristic types: V-shaped, flat-type, and peak-type.
Let us consider the Fermi surface to be an isotropic sphere.
In the Balian-Werthamer (BW) state $[\bm{d}=(\hat{k}_x,\hat{k}_y,\hat{k}_z)]$, which is the $p$-wave pairing state with a full gap, the surface hosts a Dirac cone, resulting in a V-shaped DOS, i.e., $\rho(E)\propto |E|$, near zero energy.
The chiral state, $[\bm{d}=(0,0,\hat{k}_x+i\hat{k}_y)]$, and the helical state, $[\bm{d}=(\hat{k}_x,\hat{k}_y,0)]$, have point nodes in their bulk. These states have zero-energy ABSs on a specific surface orientation. 
These ABSs are dispersionless and connect the point nodes projected onto the surface Brillouin zone, known as Fermi arcs, while dispersing along the direction perpendicular to the nodal direction. 
The large number of zero-energy states in the chiral and helical states results in a flat surface DOS, i.e., $\rho(E)= {\rm const.}$, around zero energy.
However, ABSs are absent on surfaces in the nodal directions, and the surface DOS exhibits a V-shaped profile, reflecting the nodal gap structure in their bulk.
For the polar $p_z$-wave state $[\bm{d}=(\hat{k}_z,0,0)]$ with a bulk nodal line, zero-energy ABSs appear as a two-dimensional (2D) sheet enclosed by the nodal line on the $(001)$ surface Brillouin zone, a structure sometimes referred to as a drumhead state.
This results in a prominent zero-energy peak in the surface DOS.
Again, on surfaces without ABSs, the DOS returns to a V-shaped profile.
A similar behavior is observed in the $d_{x^2-y^2}$-wave superconductor ($\psi = k_x^2-k_y^2)$, which also hosts line nodes.
On the $(110)$ surface, drumhead-like ABSs emerge, producing a sharp zero-energy peak in the surface DOS.

\section{Tunneling current and shot noise in superconducting STM\label{Sec4}}
In this section, we calculate the $dI/dV$ spectra and current noise for various types of superconducting junctions,
as discussed in the previous section, in order to simulate STS measurements with an $s$-wave superconducting tip.
The tunneling current is evaluated from Eq.~\eqref{eq:Ifull} by solving the Dyson equation~\eqref{eq:Dyson_GLR} for the momentum-averaged Green’s function, which is appropriate for local tunneling processes where momentum is not conserved.
The current noise is calculated from Eqs.~\eqref{eq:Noisefull} and \eqref{eq:IIcorr}, and the corresponding Fano factor is obtained to simulate shot-noise spectroscopy.
These formulations naturally incorporate tunneling processes to all orders.
In the weak-tunneling regime, the characteristic features of $dI/dV$ spectra and the current noise can be understood analytically from Eqs.~\eqref{eq:Isingle} and \eqref{eq:IAndreev} derived in Sec.~\ref{2A}.

We assume that the left electrode is an $s$-wave superconducting tip,
whose retarded Green's function is given by
\begin{align}
\label{eq:Green_s}
    G^r_L(\omega) = \frac{-\pi N_L}{\sqrt{\Delta_L^2-(\hbar\omega+i\delta_L)^2}}\begin{pmatrix}
        \hbar\omega+i\delta_L & \hat{\Delta}_L \\
        \hat{\Delta}_L & \hbar\omega+i\delta_L
    \end{pmatrix}, 
\end{align}
where the gap function is
\begin{eqnarray}
    \hat{\Delta}_L = \begin{pmatrix}
        & \Delta_L \\
        -\Delta_L &
    \end{pmatrix}.
\end{eqnarray}
The DOS is obtained from the retarded Green's function as
\begin{eqnarray}
    \rho_L(E) &=& -\frac{i}{\pi}\Im\mathrm{Tr}_{e}[G_L^r(E/\hbar)],  
\end{eqnarray}
where $\mathrm{Tr}_{e}$ denotes the trace over the electron (normal) part of the Green's function.
The parameter $\delta_L$, referred to as the Dynes parameter, represents a phenomenological broadening factor~\cite{dynes1978direct}.
Microscopically, it arises from processes such as impurities or phonon scattering and accounts for the finite lifetime of quasiparticle states.
In this work, we set  $\delta_L/\Delta_L = 0.01$, which is a typical value of recent experiments with clean tips~\cite{gu2023detection,esat2023determining}.
The experimental temperature is taken to be $\Theta/\Theta_{Lc}=0.05$,
where $\Theta_{Lc}$ is the superconducting transition temperature of the left electrode, defined as $\Theta_{Lc}=\Delta_L/1.76$.

For the right electrode, we consider representative topological superconductors listed in Table~\ref{list_SC}.
The momentum-averaged Green's function is defined by
\begin{eqnarray}
    G_R^r(\omega) = \sum_{\bm{k}}G_R^r(\bm{k},\omega),
\end{eqnarray}
where $G^r_R(\bm{k},\omega)$ is obtained from the tight-binding model Hamiltonian using the recursive Green's function method~\cite{umerski1997closed,ohashi2024anisotropic}.
The normal-state Hamiltonian is provided in Appendix~\ref{Appendix:model}.
One might expect that momentum averaging in anisotropic superconductors eliminates information about anomalous pair correlations.
However, due to the breaking of spatial inversion symmetry at the interface, odd-frequency components arise, whose contribution remains finite even after momentum averaging~\cite{tanaka2007odd}.
The static phase difference between the two superconductors is set to zero.
As can be seen from Eqs.~\eqref{eq:Isingle} and \eqref{eq:IAndreev}, however, the dc tunneling current is independent of the superconducting phase difference.

\subsection{$dI/dV$ characteristics probed by STS with a superconducting tip}
\label{5A}

Before discussing each pairing state in detail, let us first clarify the characteristic features of the differential conductance spectrum obtained in STS measurements with an 
$s$-wave superconducting tip.
Since STM typically operates in the weak-coupling regime~($\alpha\ll1$), the overall behavior of $dI/dV$ can be captured by the fourth-order tunneling processes derived in Sec.~\ref{2A}.
From Eqs.~\eqref{eq:Isingle} and \eqref{eq:IAndreev}, these currents reduce to simple expressions at zero temperature and in the low-bias limit,
\begin{gather}
    \label{eq:Isingle2}
    I_{\rm qp}(V) \propto \int^{eV}_{0}dE~\rho_L(E-eV)\rho_R(E), \\
    \label{eq:Andreev2}
    I_{\rm AR}(V) \propto \int_{-eV}^{eV}dE~\rho_R(-E+eV)\rho_R(E+eV).
\end{gather}
In this subsection, we use these expressions to discuss the characteristic peak structures in the $dI/dV$ spectra.
For simplicity, we consider the case of $V>0$.

We begin by discussing the characteristic spectral shape of the single-particle tunneling contribution, $dI_{\rm qp}/dV$.
From Eq.~\eqref{eq:Isingle2}, the $dI/dV$ for this process is given by
\begin{eqnarray}
    \dv{I_{\rm qp}}{V} \propto \int_0^{eV}dE~\rho_L'(E-eV)\rho_R(E)
\end{eqnarray}
where $\rho'(E)$ denotes the derivative of the quasiparticle DOS with respect to energy.
Here, we have used the fact that $\rho_L(0) = 0$ for an $s$-wave superconductor.
Since the quasiparticle DOS in an $s$-wave superconductor vanishes for $E<\Delta_L$ due to the superconducting gap, we approximate its derivative by a delta function, $\rho_L(E)\sim\delta(E\pm\Delta_L)$. 
Under this approximation,
\begin{eqnarray}
    \label{eq:dIqp/dV}
    \dv{I_{\rm qp}}{V} \propto -e\rho_R(eV-\Delta_L)\theta(eV-\Delta_L),
\end{eqnarray}
where $\theta(x)$ is the step function.
Thus, the lowest voltage at which the single-particle process starts to contribute is $eV=\Delta_L$.
Furthermore, if the DOS in the topological superconductor has a coherence peak at $E=\Delta_R$, the factor $\rho_R$ appearing in Eq.~\eqref{eq:dIqp/dV} becomes maximal at $eV = \Delta_L + \Delta_R$.
This feature is known as the total coherence peak in the case of junctions of $s$ -wave /$s$ -wave.
Physically, once $eV>\Delta_L$, quasiparticle excitations become available in the $s$-wave superconductor, and the single-particle tunneling dominates.
In particular, at $eV=\Delta_L+\Delta_R$, quasiparticles tunnel between the coherence peak of the topological superconductor and that of the $s$-wave superconductor, leading to a pronounced peak in the $dI/dV$ spectrum. 
Therefore, single-particle tunneling starts to contribute at $V=\Delta_{L}$, and the spectrum of $dI_{\rm qp}/dV$ may exhibit sharp peaks at $eV=\Delta_{L}$ and $eV=\Delta_L+\Delta_R$.
In the low-bias regime $eV < \Delta_L$, the contribution from single-particle tunneling vanishes, and the $dI/dV$ characteristic is therefore dominated by the Andreev reflection process.
From Eq.~\eqref{eq:Andreev2}, the $dI/dV$ associated with the Andreev reflection is given by
\begin{align}
    \label{eq:dIdV_AR}
    \dv{I_{\rm AR}}{V}\propto& ~\rho_R(2eV)\rho_R(0)\nn \\
    & -2\int_0^{2eV}dE~\rho_R'(E-2eV)\rho_R(E),
\end{align}
where the particle-hole symmetry $\rho_R(E)=\rho_R(-E)$ is assumed.
If Majorana zero modes are present and induce a finite zero-energy DOS $(\rho_R(0)\neq 0)$, the first term gives rise to a subgap peak structure at $eV = \Delta_R/2$.
Furthermore, by approximating the derivative of $\rho_R(E)$ as $\delta(E\pm \Delta_R)$, one finds that the second term yields a pronounced contribution at $eV=\Delta_R$.

We conclude that, in Andreev spectroscopy on a topological superconductor that exhibits coherence peaks at $E=\pm\Delta_{R}$, one can find characteristic conductance peaks at
$eV = \pm\Delta_R,~\pm\Delta_L,~\pm(\Delta_R + \Delta_L)$.
Additionally, if the topological superconductor hosts finite zero-energy DOS $(\rho_R(0)\neq 0)$,
a peak may also appear at $eV = \pm \Delta_R/2$.
It is worth noting the appearance of negative differential conductance $dI/dV < 0$.
In S-S junctions, both electrodes can exhibit singularities in the DOS, which resonate at certain bias voltages and lead to a nonmonotonic increase of the tunneling current.
In particular, once the bias moves away from resonance, the tunneling probability decreases more rapidly than the increase in available tunneling carriers, resulting in negative $dI/dV$.
This phenomenon is absent for a normal metallic tip, whose DOS is nearly energy independent.

We remark on the possible influence of the Josephson current on the $dI/dV$ spectra in junctions with a finite Josephson coupling.
In STM-like junctions with small capacitive coupling, the Josephson current experiences a substantial change in Coulomb energy during tunneling.
Consequently, the transport enters the regime of dynamical Coulomb blockade, in which coupling to the environmental impedance gives rise to inelastic Cooper-pair tunneling.
These inelastic processes manifest themselves as a zero-bias peak (ZBP) in the $dI/dV$ spectra~\cite{ingold1994cooper}.

\subsection{$s$-wave SC\label{5B}}
\begin{figure*}[t]
    \centering
    \includegraphics[width=\linewidth]{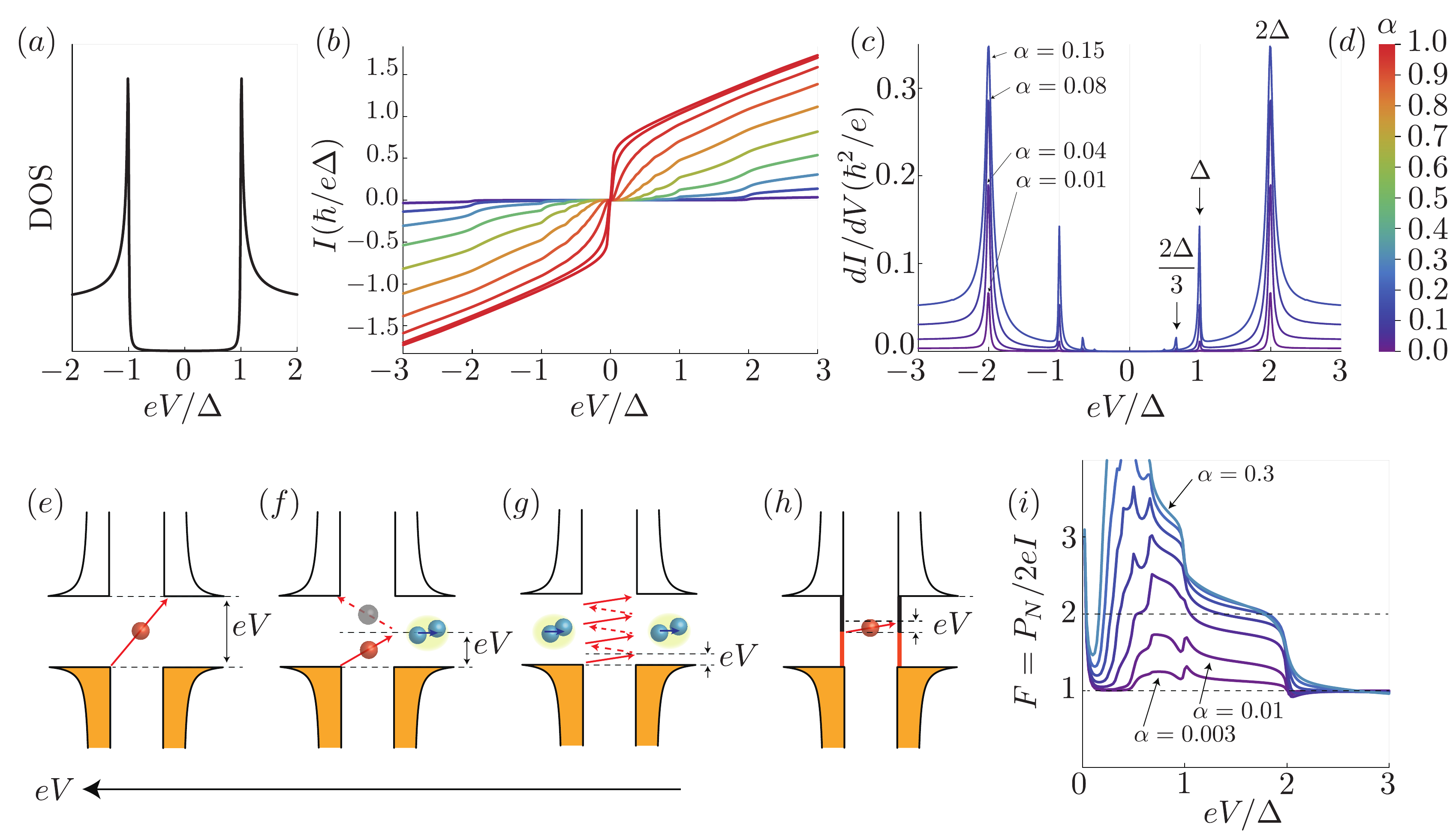} 
    \caption{
    (a)~The DOS of an $s$-wave superconductor with the Dynes factor $\delta = 0.01\Delta$.
    (b)~Current-voltage ($I$-$V$) characteristics of an $s$-wave/$s$-wave superconducting junction for various effective transparencies $\alpha$.
    The corresponding values of $\alpha$ are indicated by the color bar in panel~(d).
    (c)~Differential conductance $dI/dV$ in the weak tunneling regime.
    Subharmonic gap structures due to MAR appear at $|eV| = 2\Delta/n$.
    (e)~Single-particle tunneling process at $|eV|=2\Delta$.
    (f)~Single Andreev reflection at $|eV| = \Delta$.
    (g)~MAR at sub-gap bias.
    (h)~Single particle tunneling mediated by residual states inside the gap at low bias. 
    (i)~Fano factor $F=P_N/2eI$.}
    \label{fig:swave}
\end{figure*}

\begin{figure*}[t]
    \centering
    \includegraphics[width=\linewidth]{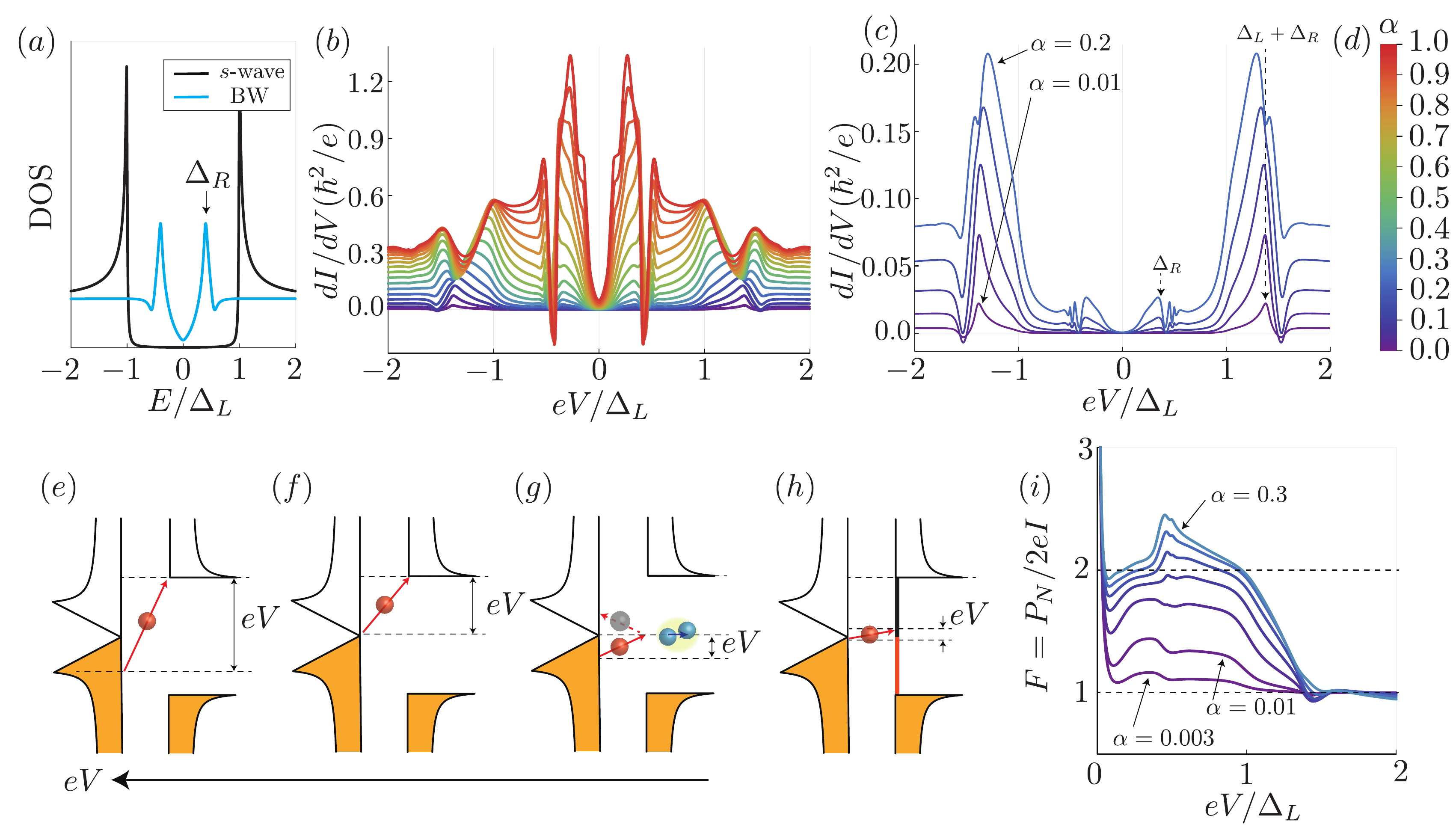} 
    \caption{Probing the {$p$-wave} BW state (right electrode) using an $s$-wave superconducting STM tip (left electrode).
    (a)~Surface DOS of the BW state and $s$-wave superconductor,
    with the Dyens parameters $\delta_L=0.01\Delta_L$ and $\delta_R = 0.025\Delta_L$.
    (b)~The $dI/dV$ spectra for various effective transparencies $\alpha$.
    The corresponding values of $\alpha$ are indicated by the color bar in panel~(d).
    (c)~The $dI/dV$ spectra in the weak tunnel regime.
    (e)~Single-particle tunneling between the coherence peak of the BW state and one of the $s$-wave superconductors.
    (f)~Single particle tunneling at $eV = \Delta_L$.
    (g)~Single Andreev reflection.
    (h)~Single particle tunneling mediated by residual states inside the gap at low bias. 
    (i)~Fano factor $F=P_N/2eI$.}
    \label{fig:BW}
\end{figure*}

\begin{figure*}[t]
    \centering
    \includegraphics[width=\linewidth]{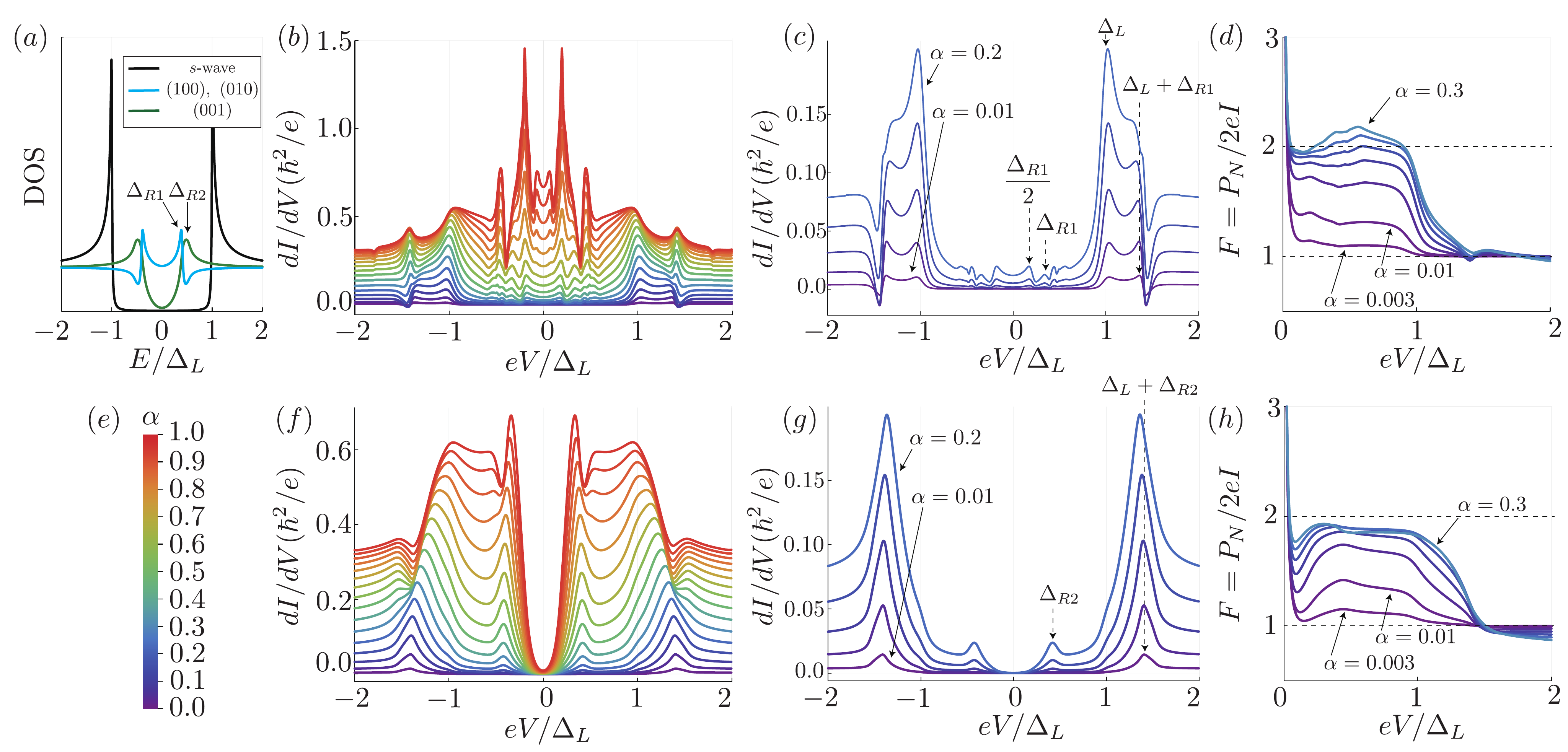} 
    \caption{Probing the chiral or helical state (right electrode) using an $s$-wave superconducting STM tip (left electrode).
    (a)~Surface DOS of the chiral superconductor for the $(100)$, $(010)$, and $(001)$ planes, together with that of the $s$-wave superconductor. The Dynes parameters are set to $\delta_R = 0.025\Delta_L$ and $\delta_L = 0.01\Delta_L$.
    (b,c)~Calculated $dI/dV$ spectra for tunneling into the $(100)$ and $(010)$ surfaces of the chiral state, for different values of the transparency $\alpha$.
    The values of $\alpha$ are indicated by the color bar in panel~(e).
    (d)~Corresponding Fano factor.
    (f,g)~Calculated $dI/dV$ spectra for tunneling into the $(001)$ surfaces of the chiral state.
    (h)~Corresponding Fano factor.}
    \label{fig:chiral}
\end{figure*}

\begin{figure*}[t]
    \centering
    \includegraphics[width=\linewidth]{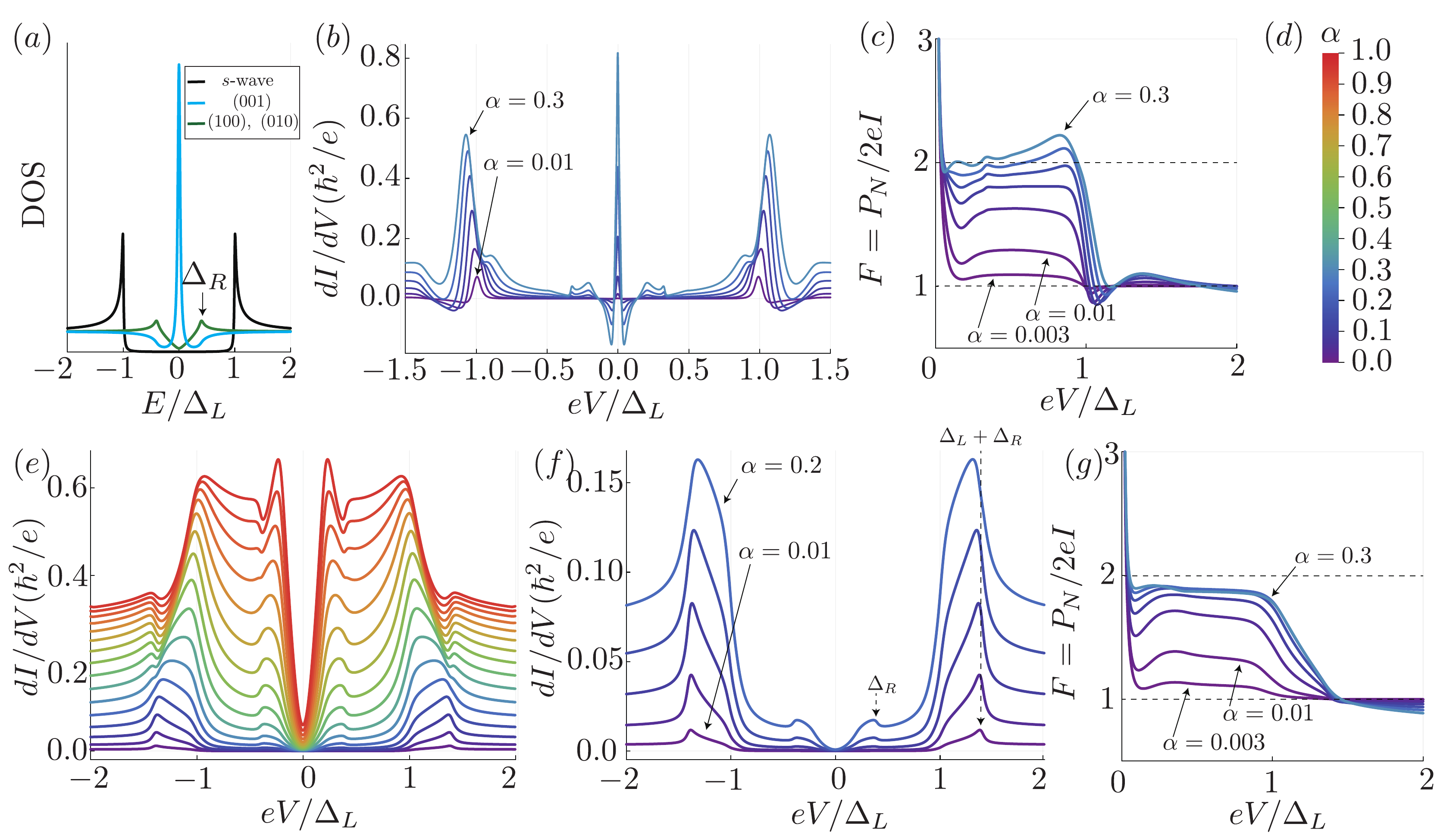} 
    \caption{Probing the polar state (right electrode) using an $s$-wave superconducting STM tip (left electrode).
    (a)~Surface DOS of the chiral superconductor for the $(100)$, $(010)$, and $(001)$ planes, together with that of the $s$-wave superconductor. The Dynes parameters are set to $\delta_R = 0.025\Delta_L$ and $\delta_L = 0.01\Delta_L$.
    (b)~Calculated $dI/dV$ spectra for tunneling into the $(001)$ surface in the weak tunneling regime $\alpha\ll 1$.
    The transparencies $\alpha$ are indicated by the color bar in panel~(d).
    (c)~Corresponding Fano factor.
    (e,f)~Calculated $dI/dV$ spectra for tunneling into the $(100)$ and $(010)$ surfaces: (e) $dI/dV$ spectra for $\alpha \in [0,1]$ and (f) for $\alpha \le 0.2$.
    (g)~Corresponding Fano factor.}
    \label{fig:polar}
\end{figure*}

As a starting point, we first consider the case where both electrodes are conventional $s$-wave superconductors. 
For simplicity, we assume that the two superconductors are identical, with $\Delta_L=\Delta_R=\Delta$.
Figure~\ref{fig:swave}(a) shows the DOS, which exhibits a fully gapped structure with sharp coherence peaks at $E = \pm \Delta$.
Figure~\ref{fig:swave}(b) shows the current-voltage ($I$-$V$) characteristics for various effective transparencies $\alpha$, which are consistent with previous theoretical works~\cite{klapwijk1982explanation,octavio1983subharmonic,bratus1995theory,cuevas1996hamiltonian}.

Let us first focus on the weak tunneling regime ($\alpha \ll 1$, see Eq.~\eqref{eq:alpha}).
Figure~\ref{fig:swave}(c) displays the corresponding $dI/dV$ spectra.
Due to the presence of the superconducting gap, the tunneling current is strongly suppressed at low bias.
Several peaks appear at $eV=\pm 2\Delta,~\pm\Delta,\pm 2\Delta/3,~\dots,\pm2\Delta/n,~\dots$
These peaks can be associated with the onset of single-particle tunneling, single Andreev reflection, and $n$-th MAR at the corresponding threshold voltages, as we explain below.
The most pronounced peak occurs at $eV = \pm 2\Delta$.
Since quasiparticle excitations are gapped by $2\Delta$, single-particle tunneling becomes possible only for $|eV| > 2\Delta$, as illustrated in Fig.~\ref{fig:swave}(e).
This onset coincides with the appearance of the total coherence peak, which corresponds to tunneling between the coherence peaks of the two superconductors~(see also Sec.~\ref{5A}).
The next prominent peak at $eV = \pm\Delta$ originates from single Andreev reflection process, as depicted in Fig.~\ref{fig:swave}(f).
Through a single Andreev reflection at the interface, an incoming electron (hole) from the left electrode is converted into a hole (electron), while a Cooper pair is simultaneously transferred into (or out of) the right superconductor.
Furthermore, higher-order MAR processes give rise to additional subharmonic peaks at $eV = \pm 2\Delta/n~(n\ge 3)$~[Fig.~\ref{fig:swave}(g)].
In the strong tunneling regime, near perfect transparency ($\alpha = 1$), 
MAR processes of arbitrarily high order contribute to the current.
In the limit $n \to \infty$, such processes can occur even at zero bias, leading to a finite tunneling current at zero bias voltage.
Consequently, $dI/dV$ diverges at zero bias~\cite{cuevas1996hamiltonian}.

The nature of the tunneling processes, including single particle tunneling, Andreev reflection, and MAR, is directly reflected in the bias dependence of the current noise.
Figure~\ref{fig:swave}(i) shows the numerically calculated Fano factor for $\alpha\ll 1$, taking into account tunneling processes to all orders.
For $|eV| > 2\Delta$, the Fano factor approaches $F=1$, reflecting the fact that single-particle tunneling dominates the transport across the junction.
In the intermediate regime $\Delta < |eV| < 2\Delta$, single-particle tunneling is suppressed, and Andreev reflection becomes the dominant process.
In the bias window $2\Delta/(n+1) < |eV| < 2\Delta/n$, the $n$th-order MAR process provides the leading contribution to the current, giving rise to a characteristic step-like structure in the Fano factor as the bias voltage is reduced~\cite{cuevas1999shot,cuevas2003full}.
At still lower bias, however, low-order tunneling processes mediated by the residual DOS, as shown in Fig.~\ref{fig:swave}(h), compete with higher-order MAR processes.
As a consequence, the $F$–$V$ curve evolves from the step-like structure and gradually approaches $F=1$ as the bias is further reduced.
At even lower bias, $|eV|<k_{\rm B}\Theta$, with $k_{\rm B}\Theta = 0.05\Delta$, thermal noise dominates the current fluctuations, leading to a divergence of the Fano factor.

Furthermore, tunneling processes mediated by the residual DOS cause the Fano factor to remain close to $F=1$ over a wide bias range in the weak-tunneling limit $\alpha \ll 1$.
Specifically, although the single-particle contribution is proportional to the Dynes parameters $\delta_{L,R}$ in the Green’s functions and is therefore intrinsically small, it scales as $|T|^2$ and consequently dominates over the $|T|^4$ contribution from Andreev reflection in the limit $\alpha \ll 1$.
As a result, in Andreev spectroscopy using superconducting tips, it is crucial that the tip is clean and characterized by a sufficiently small Dynes parameter.
Moreover, even within the weak-tunneling regime, the tunneling amplitude must be sufficiently large so that the $|T|^4$ contribution from single Andreev reflection is not overwhelmed by the $|T|^2$ single-particle tunneling process mediated by the residual DOS.
This consideration is consistent with noise measurements in mesoscopic superconducting junctions, where an effective charge of $e^* = e$ rather than $e^*=2e$ has been observed~\cite{niu2024why}.

\subsection{$p$-wave BW state}
Let us consider the BW $p$-wave state as a superconducting sample, where the gap function is given by
\begin{eqnarray}
    \hat{\Delta}_R = \Delta_{\rm BW}\begin{pmatrix}
        -\sin k_x + i\sin k_y & \sin k_z \\
        \sin k_z & \sin k_x+i\sin k_y 
    \end{pmatrix},
\end{eqnarray}
where $\Delta_{\rm BW}$ denotes the gap magnitude.
We employ a tight-binding model for the right electrode, and the corresponding normal-state Hamiltonian is given in Appendix~\ref{Appendix:model}.
The BW state serves as a prototypical example of a topological superconductor.
The BW state is established as the ground state of the B-phase of superfluid $^3$He~\cite{bw,leg75,vollhardt,volovik,miz16}. 
It has also been proposed to be realized in the superconductor UTe$_2$~\cite{matsumura2023large,suetsugu2024fully}.

Figure~\ref{fig:BW}(a) shows the surface DOS with the Dynes parameter set to $\delta_R = 0.025\Delta_L$.
All energies are expressed in units of the $s$-wave superconducting gap $\Delta_L$.
Throughout the calculations, we assume that the superconducting gap for the BW state is smaller than that for the $s$-wave superconductor, that is, $2.5\Delta_{\rm BW} = \Delta_L$.
The bulk of the BW state is fully gapped and characterized by a nontrivial 3D winding number~\cite{volovik,sch08,miz16}.
As a consequence of the bulk-boundary correspondence, linearly dispersing energy bands emerge at the boundary.
The surface DOS exhibits a V-shaped structure due to the contribution from these 2D Majorana cones.
In addition, the surface DOS exhibits coherence peaks at finite energy.
We refer to the energy position of these coherence peaks as $E=\pm \Delta_R$.
It should be noted that on the surface, the formation of ABSs tends to shift the coherence peaks to energies inside the bulk gap $\Delta_{\rm BW}$, {i.e., $\Delta_{R}<\Delta_{\rm BW}$}.
As a result, the position of the coherence peaks in the surface DOS depends on the presence or absence of ABSs and may vary between different surfaces.

Figure~\ref{fig:BW}(b) shows the calculated $dI/dV$ spectra for various effective transparencies, while Fig.~\ref{fig:BW}(c) represents the result for the weak tunneling regime.
We begin our discussion with the spectrum shown in Fig.~\ref{fig:BW}(c), which is relevant to the regime of STM experiments.
As discussed in Sec.~\ref{5A}, several characteristic peaks appear around $eV = \pm \Delta_R, \pm(\Delta_L + \Delta_R)$.
For $|eV|>\Delta_L$, the dominant contribution arises from single-particle tunneling, illustrated in Fig.~\ref{fig:BW}(f), leading to a significant increase in the tunneling current.
In particular, a total coherence peak emerges at $eV = \pm ( \Delta_L + \Delta_R)$, corresponding to single-particle tunneling from the coherence peak of the BW state to that of the $s$-wave superconductor.
Slightly above this resonance bias, $dI/dV$ may take negative values due to the rapid decrease of the single-particle tunneling probability.
For $|eV|<\Delta_L$, the tunneling process via Andreev reflection becomes dominant and produces characteristic subpeak structures, including a peak at $eV=\Delta_R$ reflecting the coherence peak of the DOS. 
In the zero-bias limit, the linear DOS yields a super-Ohmic behavior with $I_{\rm AR}\propto V^3$.
Since the total coherence peak arises from a second-order (${|T|}^2$) process, whereas the low-bias Andreev contribution originates from a fourth-order (${|T|}^4$) process, the latter becomes more pronounced as the transparency, $\alpha$, increases.
In the intermediate and strong tunneling regimes, the subpeak at $eV=\pm \Delta_R$ develops prominently and can even exceed the total coherence peak near $\alpha\to 1$.
Nevertheless, unlike in the $s$-wave/$s$-wave case, the coherent zero-bias current does not occur, since ordinary backscattering processes coexist with MAR.
It should be remarked that the expression for $\alpha$ of Eq.~\eqref{eq:alpha} is strictly valid only for $s$-wave/$s$-wave junctions. In other cases, it should be regarded merely as an indicator of the tunneling strength.

Let us discuss the noise properties of the tunneling current.
Figure~\ref{fig:BW}(i) shows the Fano factor calculated from Eqs.~\eqref{eq:Noisefull} and \eqref{eq:IIcorr}.
For bias voltage above the total coherence peak, $|eV|>\Delta_L+\Delta_R$, the single-particle tunneling process dominates, and the Fano factor approaches $F=1$.
In the intermediate range $\Delta_L<|eV|<\Delta_L+\Delta_R$, the single-particle tunneling process is gradually suppressed [Fig.~\ref{fig:BW}(f)],
and the relative contribution of the Andreev reflection process increases, leading to an enhanced Fano factor.
For $|eV|<\Delta_L$, the single particle tunnel is completely suppressed, and Andreev reflection becomes the dominant process, causing the Fano factor to approach $F=2$.
It should be noted that, unlike in conventional $s$-wave/$s$-wave junctions, single Andreev reflection provides the leading contribution in the regime $|eV| < \Delta_L$.
As a consequence, the Fano factor does not significantly exceed $F=2$, and no step-like structure appears in the $F$–$V$ curve.
In the zero bias limit ($|eV|<k_{\rm B}\Theta$), thermal noise dominates, leading to a divergence of the Fano factor.

In the weak tunneling regime $\alpha \ll 1$, due to the finite residual DOS in the $s$-wave superconductor, as illustrated in Fig.~\ref{fig:BW}(h), single-particle tunneling remains possible and competes with Andreev reflection, preventing $F$ from reaching $F=2$.
In particular, reflecting the super-Ohmic behavior of the Andreev current, the Fano factor returns toward $F=1$ in the regime $|eV|<\Delta_R$.

\subsection{$p$-wave chiral/helical state}
As a second example, we consider $p$-wave chiral and helical superconducting states, often referred to as Weyl and Dirac superconductors, respectively.
In a spin-full system, the gap function of the chiral state takes the form
\begin{eqnarray}
    \label{eq:chiral}
    \hat{\Delta}_R = \Delta_{\rm C}\begin{pmatrix}
        0 & \sin k_x + i\sin k_y \\
        \sin k_x + i\sin k_y & 0
    \end{pmatrix}.
\end{eqnarray}
The bulk spectrum hosts point nodes at $k_x = k_y = 0$, around which quasiparticles acquire a nontrivial Berry phase.
A 2D momentum slice between the nodes realizes an effective quantum Hall state characterized by a nontrivial Chern number, leading to chiral edge states.
As a result, 1D dispersionless ABSs---Fermi arcs---emerge, terminating at the projections of the bulk nodes on the surface Brillouin zone.
The chiral pairing state is realized in the A phase of superfluid $^3$He~\cite{and61,and73,leg75,vollhardt,volovik,miz16} and has been proposed for uranium-based ferromagnetic superconductors~\cite{aoki2001coexistence,huy2007superconducitivty}.

A helical superconductor can be constructed as a superposition of chiral states with opposite chirality, described by the gap function
\begin{eqnarray}
    \label{eq:helical}
    \hat{\Delta}_R = \Delta_{\rm H}\begin{pmatrix}
        -\sin k_x + i\sin k_y & 0 \\
        0 & \sin k_x + i\sin k_y
    \end{pmatrix}.
\end{eqnarray}
In this case, the time-reversal symmetry is restored.
The candidate pairing states of UTe$_2$---$B_{1u}$, $B_{2u}$, and $B_{3u}$ states, correspond to the helical pairing states~\cite{gu2025pair,wang2025imaging,theuss2024single,lee2025anisotropic,hayes2025robust,zixuan2025obserbation}.

The essential distinction between chiral and helical superconductors lies in the presence or absence of time-reversal symmetry.
From the viewpoint of electrical probes, such as tunneling current, however, the two are difficult to distinguish.
Indeed, the DOS calculated from the gap function in Eqs.~\eqref{eq:chiral} and \eqref{eq:helical} are identical, and the resulting dc tunneling current is likewise the same.
In what follows, we therefore focus on the chiral case, while noting that the dc transport properties are equivalent for the helical state.
We remark, however, that the corresponding ac response may differ between the two.

Figure~\ref{fig:chiral}(a) shows the surface DOS for the chiral state, where the gap magnitude and the Dynes parameter are set to $2.5\Delta_C = \Delta_L$ and $\delta_R = 0.025\Delta_L$, respectively.
In this model, the $(100)$ and $(010)$ surfaces are equivalent; in the following, we therefore focus on the $(100)$ and $(001)$ surfaces.
For the gap function in Eq.~\eqref{eq:chiral}, Fermi arcs appear on the $(100)$ surface.
As a result, the DOS exhibits a finite zero-energy value with a flat-type energy dependence, together with a coherence peak at $E=\pm\Delta_{R1}$.
By contrast, no surface states exist on the $(001)$ surface; instead, the DOS displays a V-shaped (or U-shaped) spectrum reflecting the bulk point nodes, with a coherence peak at $E=\pm\Delta_{R2}$.
Due to the formation of ABSs on the $(100)$ surface, the coherence peak appears at lower energy than on the $(001)$ surface, i.e., $\Delta_{R1} < \Delta_{R2}$.

We first discuss the (100) surface.
Figure~\ref{fig:chiral}(b) shows the calculated $dI/dV$ spectra for various effective transparencies, while Fig.~\ref{fig:chiral}(c) represents the result for the weak tunneling regime.
In the weak tunneling regime, several peaks appear at $eV = \pm\Delta_{R1}/2,~\pm\Delta_{R1},~\pm\Delta_L$, and $\pm(\Delta_L+\Delta_{R1})$
(see Sec.~\ref{5A}).
The peak at $eV=\pm(\Delta_L+\Delta_{R1})$ corresponds to the total coherence peak, as illustrated in Fig.~\ref{fig:BW}(e).
Since $eV=\Delta_L$ marks the onset of single-particle tunneling [Fig.~\ref{fig:BW}(f)], the finite zero-energy DOS [$\rho_{R}(0)\neq 0$] leads to a pronounced peak at this bias, reflecting the coherence peak of the $s$-wave electrode.
As discussed in Eq.~\eqref{eq:Andreev2}, the coherence peak at $E=\pm\Delta_{R1}$ of the chiral state is transferred via Andreev reflection processes to features at $eV=\pm\Delta_{R1}/2$ and $eV=\pm\Delta_{R1}$.
In particular, the peak at $eV=\pm\Delta_{R1}/2$ arises due to the finite zero-energy DOS.
Toward the strong-tunneling regime, the peak at $eV=\pm\Delta_{R1}$ becomes most pronounced, similar to the BW case.
Figure~\ref{fig:chiral}(d) shows the corresponding Fano factor.
For $|eV|>\Delta_L$, single-particle tunneling dominates, and the Fano factor approaches $F=1$.
For $|eV|<\Delta_L$, single-particle tunneling is completely suppressed, and Andreev reflection dominates, leading to $F$ approaching $F=2$.
However, due to the finite residual DOS, single-particle tunneling persists, and for $\alpha \ll 1$, the Fano does not reach $F=2$.

We next turn to the $(001)$ surface.
Since its surface DOS closely resembles that of the BW state, the resulting $dI/dV$ spectra and noise characteristics are nearly identical.
We therefore avoid repeating the discussion.
This demonstrates that the dc tunneling properties probed by STS are entirely governed by the surface DOS.

\subsection{$p$-wave polar state}
\begin{figure*}[t]
    \centering
    \includegraphics[width=\linewidth]{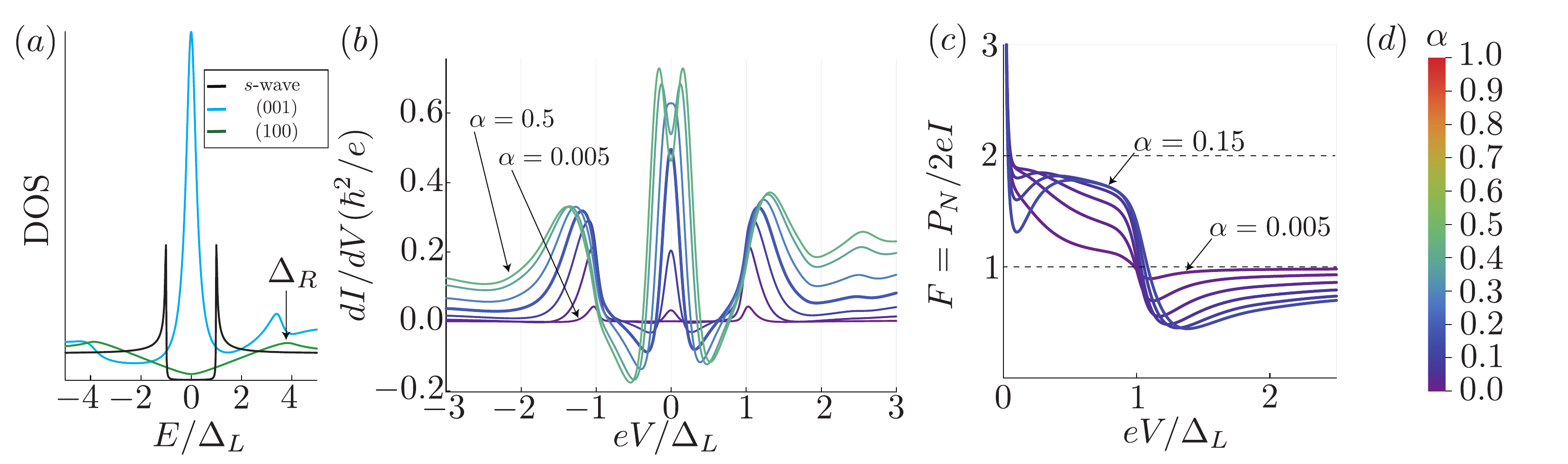} 
    \caption{Probing the $(110)$ surface of the $d_{x^2-y^2}$ state (right electrode) using an $s$-wave superconducting STM tip (left electrode).
    (a) Surface DOS of the $d_{x^2-y^2}$ state for the $(100)$ and $(110)$ orientations, together with that of the $s$-wave superconductor.
    The gap amplitude of the right electrode is taken to be about four times larger than that of the left, denoted as $\Delta_R$.
    The Dynes parameters are set to $\delta_R = 0.25\Delta_L$ and $\delta_L = 0.01\Delta_L$.
    (b) Calculated $dI/dV$ spectra for tunneling into the $(110)$ surface from the weak to intermediate transparency regimes.
    The corresponding values of $\alpha$ are indicated by the color bar in panel (d).
    (c) Fano factor corresponding to the spectra in (b).
    Calculations are performed using the standard $t$–$t'$ tight-binding model.}
    \label{fig:dwave}
\end{figure*}

We consider the $p$-wave polar state, whose gap function is given by
\begin{eqnarray}
    \hat{\Delta}_R = \Delta_{\rm P}\begin{pmatrix}
        -\sin{k_z} & 0 \\
        0 & \sin{k_z}
    \end{pmatrix}.
\end{eqnarray}
In the bulk, a line node where the quasiparticle excitation becomes gapless appears along the equator of the Fermi surface.
The polar phase has been realized in superfluid $^3$He under anisotropic nanoconfinement~\cite{aoyama,dmitriev},
whereas it has not been reported in superconductors.

Since the $(100)$ and $(010)$ surfaces are equivalent, we focus on the $(100)$ and $(001)$ surfaces.
Figure~\ref{fig:polar}(a) displays the surface DOS, where the gap magnitude and the Dynes parameter are set to $2.5\Delta_{\rm P} = \Delta_L$ and $\delta_R = 0.025\Delta_L$, respectively.
On the $(100)$ surface, no surface states appear, and the bulk line nodes give rise to a V-shaped energy dependence in the DOS.
In contrast, on the $(001)$ surface, zero-energy states emerge within the 2D area of the surface Brillouin zone enclosed by the line node, resembling drumhead-like surface states in nodal line semimetals~\cite{bian2016drumhead, yamakage2016line}.
Reflecting the presence of abundant gapless excitations, the DOS exhibits a pronounced peak at zero energy.

We first discuss the tunneling current and noise characteristics for the $(001)$ surface.
Figure~\ref{fig:polar}(b) displays the $dI/dV$ spectra in the low-transparency regime.
While single-particle tunneling generally starts at ${|eV|}=\Delta_L$, the sharp zero-energy peak in the DOS of the polar state makes this bias an enhanced  resonance point, giving rise to the total coherence peak.
Another notable feature is the pronounced zero-bias conductance peak, which originates from the sharp zero-energy DOS through the Andreev reflection process between the 2D flat band in the polar state and Cooper pairs in the $s$-wave superconductor.
Owing to the extremely large DOS in the polar state, the ZBP arising from the ${|T|}^4$ tunneling process can become comparable to the total coherence peak, even in the low-transparency regime.
As the bias voltage is shifted away from zero, the Andreev tunneling probability rapidly diminishes, giving rise to a large negative $dI/dV$.
Note that the results for high transparency are not presented, since the zero-bias structure becomes numerically unstable.

Figure~\ref{fig:polar}(c) shows the corresponding Fano factor.
For $|eV| > \Delta_L$, the current is dominated by single-particle tunneling, yielding $F = 1$, whereas for $|eV| < \Delta_L$, single-particle tunneling is strongly suppressed and Andreev tunneling dominates, resulting in $F = 2$.
When the Josephson coupling is finite, a zero-bias peak (ZBP) can also arise from the Josephson current in the dynamical Coulomb blockade regime.
Therefore, based solely on Andreev STS measurements, it is difficult to distinguish whether the ZBP originates from Andreev reflection or Josephson tunneling.
Since the Josephson current is intrinsically coherent, its noise characteristics are markedly different.
Consequently, an observed value of $F = 2$ provides direct evidence that the tunneling current originates from Andreev reflection.

For the $(100)$ surface, the DOS exhibits a V-shaped structure similar to that of the BW state.
Consequently, as shown in Figs.~\ref{fig:polar}(f-h), both the $dI/dV$ and noise characteristics closely resemble those of the BW state.

\subsection{$d_{x^2-y^2}$-wave superconductivity}
Lastly, we consider the line-nodal $d_{x^2-y^2}$ pairing as a representative example of $d$-wave superconductors, whose gap function is
\begin{align}
    \hat{\Delta}_{R} = \Delta_{d}\begin{pmatrix}
        0 & \cos(k_x)-\cos(k_y) \\
        -\cos(k_x)+\cos(k_y) & 0
    \end{pmatrix}.
\end{align}
The $d_{x^2-y^2}$ state is well established in cuprate superconductors, including YBCO and Bi-based compounds~\cite{tsuei1994pairing,scalapino1995case,tsuei2004robust}.
A characteristic feature of this state is the emergence of drumhead-like zero-energy ABSs on the $(110)$ surface, forming 2D zero-energy states within the bulk line node.
Here, we consider the $(110)$ surface, which hosts ABSs, and the $(100)$ surface, where no bound state appears.
Figure~\ref{fig:dwave}(a) shows the surface DOS for the $(110)$ surface, which exhibits a pronounced ZBP as a direct consequence of the drumhead-like zero-energy ABSs, together with the DOS for the $(100)$ surface.

Here, we discuss the $dI/dV$ behavior and noise characteristics for the $(110)$ surface.
Figure~\ref{fig:dwave}(b) shows the calculated $dI/dV$ spectra from the weak to intermediate tunneling regime.
The parameters are set to $\Delta_d = 2\Delta_L$, $\delta_R = 0.25\Delta_L$, and $\delta_L =0.01\Delta_L$.
As in the case of the polar state discussed in the previous section, the Andreev reflection tunneling processes of the drumhead-like zero-energy ABSs produce a pronounced ZBP in the $dI/dV$ spectra, while the total coherence peaks appear at $eV = \pm \Delta_L$ via single-particle tunneling processes.
In the intermediate regime of $\alpha$, however, a distinct feature emerges: when $\alpha \approx 0.4$, the ZBP splits into two peaks.
This splitting can be attributed to the sharp coherence peak of the $s$-wave electrode, which imposes a subpeak structure on the broadened ZBP.

Figure~\ref{fig:dwave}(c) shows the corresponding Fano factor.
For bias voltages above $\Delta_L$, the Fano factor remains close to, or slightly below, $F=1$, whereas for $|eV| < \Delta_L$ it increases toward $F=2$ as Andreev reflection becomes the dominant transport process.
The observation of a Fano factor $F<1$ at high bias indicates suppressed current fluctuations and signifies that the tunneling process is non-Poissonian.
This behavior originates from the large surface DOS associated with the ZBP on the $(110)$ surface, which effectively opens highly transmitting tunneling channels and suppresses current fluctuations.

\section{summary}
\label{Sec6}
Motivated by recent advances in STM with $s$-wave superconducting tips,
we have developed a general theoretical framework for Andreev and shot noise spectroscopy to probe the signatures of the surface-bound states of topological superconductors.
We have derived the effective tunneling action for a point contact between a topological superconductor and a conventional $s$-wave superconductor, including Andreev reflection processes in the real-time Keldysh formalism.
Employing the saddle-point approximation, we obtain a Langevin-type Kirchhoff current equation
that incorporates the single-particle current, the Andreev current, and their associated current noise.
In the low-bias limit, the expression for the Andreev reflection current simplifies to a form
given by the convolution of the particle and hole density of states (DOS) of the surface states
of the topological superconductor, providing a direct manifestation of the Andreev reflection process.
The analytical expressions derived for the single-particle and Andreev currents serve as good
approximations over the entire bias range in the weak-tunneling regime relevant to STM measurements.
From this analytical understanding, we find that Andreev spectroscopy exhibits characteristic
subgap peaks at $eV = \pm\Delta_R/2, \pm\Delta_R, \pm\Delta_L, \pm(\Delta_L + \Delta_R)$,
where $\Delta_L$ and $\Delta_R$ denote coherence peak energies of the left and right electrodes, respectively.
Furthermore, we have calculated the tunneling current for arbitrary transparencies by taking into account not only Andreev reflection processes but also all other tunneling processes.
Our theory is therefore applicable to experimental results from STS measurements and point-contact spectroscopy for topological superconductors in any tunneling regime.
We have also analyzed the current noise and derived an analytic expression for the Fano factor in the
shot-noise limit. 
In particular, we showed that the Fano factor associated with the Andreev current
approaches the universal value of $F=2$, reflecting the transfer of charge~$2e$ Cooper pairs.

We have systematically investigated the surface DOS, the $dI/dV$ spectra, and noise characteristics  for representative topological superconductors listed in Table~\ref{list_SC} and for their typical crystal surfaces.
The $dI/dV$ characteristics are determined by the surface DOS, which can be classified according to the dimensionality of the ABSs.
Since the appearance of ABSs is closely tied to the superconducting pairing symmetry, STS measurements of the surface DOS can serve as a decisive experimental method to directly determine the pairing symmetry.
However, conductance spectra alone cannot identify the microscopic origin of the tunneling current.
For example, a ZBP in the $dI/dV$ spectra can arise from either Josephson or Andreev tunneling.
Shot-noise measurements provide further insight, as they directly probe the microscopic mechanism of the tunneling process.
We have also demonstrated that, in the weak-tunneling limit, single-particle tunneling through the residual DOS can dominate over Andreev tunneling.
Therefore, a clean superconducting tip is essential for realizing Andreev tunneling, and the tunneling amplitude must be sufficiently large for the ${|T|}^4$ Andreev tunneling current to compete with single-particle tunneling through the residual DOS.
Our results thus serve as a theoretical benchmark for the design and fabrication of high-quality superconducting tips for STS spectroscopy.

Finally, we comment on recent experiments on the superconductor UTe$_2$~\cite{gu2025pair,wang2025imaging}.
In the STS measurements using an $s$-wave superconducting tip, a zero-bias conductance peak comparable to the total coherence peak has been observed on the naturally cleaved (011) surface.
Our analysis indicates that the emergence of such a ZBP through Andreev reflection requires the presence of a zero-energy peak in the surface DOS, suggesting the possible realization of exotic surface Majorana states on the $(011)$ surface of UTe$_2$~\cite{tei2025topological}.

\section*{ACKNOWLEDGMENTS}
J.T. is supported by a Japan Society for the Promotion of Science (JSPS) Fellowship for Young Scientists.
This work was supported by JSPS KAKENHI (Grant No.~JP23K20828, No.~JP23K22492, No.~JP24KJ1621, No.~JP25H00599, No.~JP25H00609, No.~JP25K07227, 
{No.
JP25K00935}
and No.~JP25K22011) 
and a Grant-in-Aid for Transformative Research Areas (A) ``Correlation Design Science'' (Grant No.~JP25H01250) from JSPS of Japan.

\appendix

\section{Model of superconductors\label{Appendix:model}}
\begin{figure}[t]
    \centering
    \includegraphics[width=\linewidth]{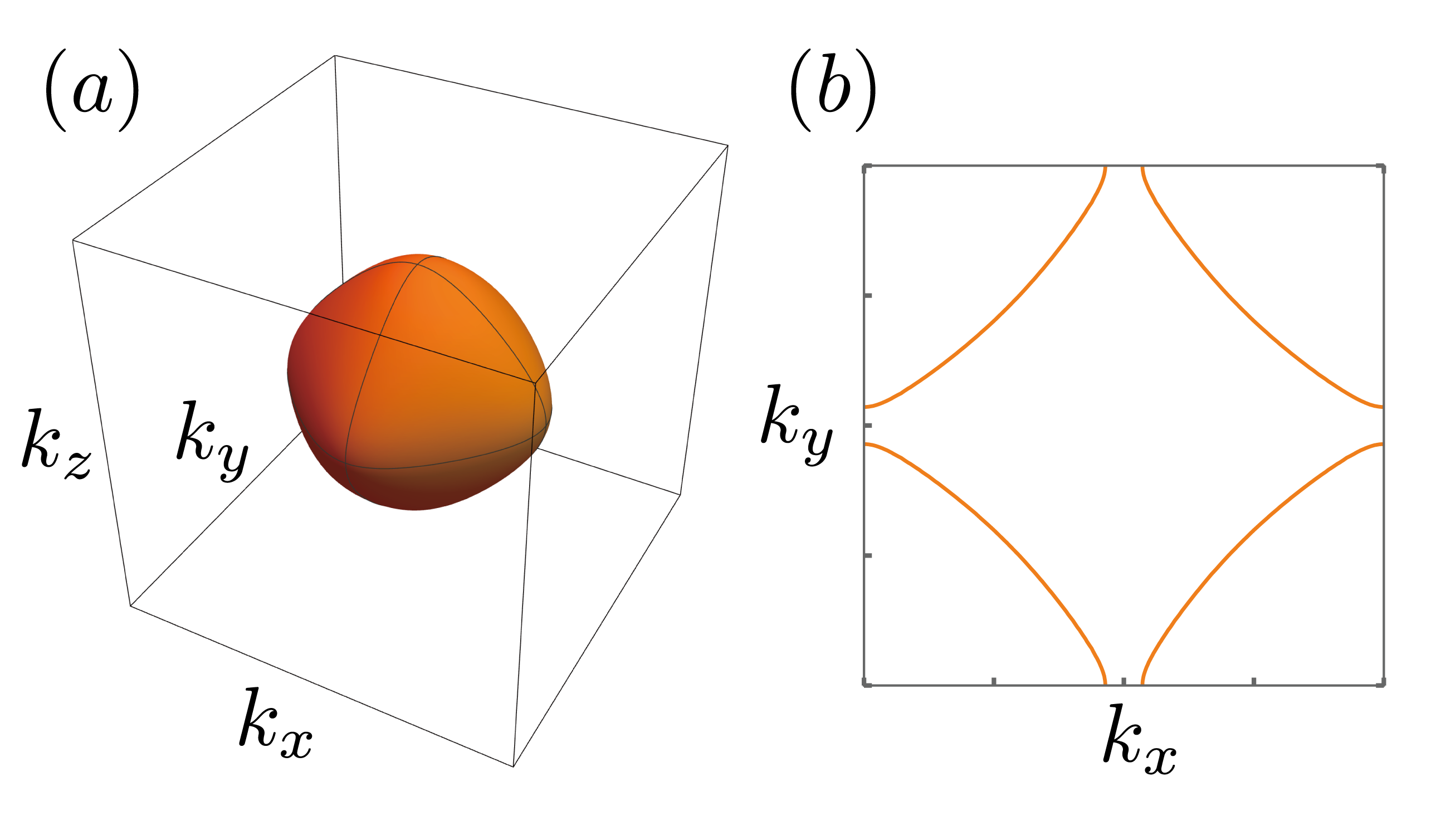} 
    \caption{Fermi surfaces for (a)~all $p$-wave superconductors and for (b)~the $d_{x^2-y^2}$ superconductor.}
    \label{fig:Fermi}
\end{figure}

The topological superconductors are modeled by a tight-binding Hamiltonian.
For the normal-state Hamiltonian of all $p$-wave superconductors, we employ
\begin{align}
    H_{N} = 2t(\cos k_x + \cos k_y + \cos k_z) - \mu,
\end{align}
where the hopping and chemical potential are set to $t/\Delta_L = -500$ and $\mu/\Delta_L = -1500$, respectively.
The corresponding Fermi surface is shown in Fig.~\ref{fig:Fermi}(a).

For the $d_{x^2 - y^2}$-wave superconductor, we adopt the standard $t$–$t'$ tight-binding model,
\begin{align}
    H_{N} = 2t_1(\cos k_x + \cos k_y) + 4t_2\cos k_x\cos k_y - \mu,
\end{align}
where the parameters are set to $t_1/\Delta_L = -25$,~$t_2/\Delta_L = 4.2$ and $\mu/\Delta_L = -15$.
The resulting Fermi surface is shown in Fig.~\ref{fig:Fermi}(b).

\section{Derivation of the effective action in Eq.~\eqref{eq:Action}\label{Appendix1}}
In this Appendix, we derive the effective action in Eq.~\eqref{eq:Action} for the phase dynamics from the microscopic Hamiltonian given in Eq.~\eqref{eq:total}.
As a first step, we introduce appropriate bosonic fields via the Hubbard-Stratonovich transformations.
The pairing interaction in Eq.~\eqref{eq:HL} is transformed by introducing complex bosonic fields $\Delta$ and $\bar{\Delta}$, which represent the superconducting order parameter:
\begin{eqnarray}
    {\mathcal{H}}_{\rm pair}'  &=& \int dr~\Big[\bar{\Delta}\psi_\downarrow\psi_{\uparrow} + \Delta{\psi}^{\dagger}_{\uparrow}{\psi}^{\dagger}_{\downarrow}
    + \frac{1}{g}|\Delta|^2 \Big],
\end{eqnarray}
where, for simplicity, we only show the case of the $s$-wave pairing here. 
The same formulation applies to anisotropic superconductors as well, although the notation becomes more cumbersome.
Next, we transform the Coulomb interaction by introducing a real voltage field $V$,
which couples to charge imbalance between the left and right electrodes:
\begin{eqnarray}
     {\mathcal{H}}_Q'= \frac{1}{2}eV(N_L-N_R) - \frac{1}{2}CV^2.
\end{eqnarray}
The resulting Hamiltonian is quadratic in the fermionic fields and can thus be integrated out by performing a Gaussian integration.
Consequently, the effective action for the bosonic fields in the Keldysh formalism is given by
\begin{eqnarray}
    S_{\rm eff} &=& -i\hbar~\mathrm{Tr} \ln \hat{G}^{-1} - \int_{{\mathcal{C}}}dx~\frac{1}{g}|\Delta_L|^2  \nn \\
    &&- \int_{{\mathcal{C}}}dx~\frac{1}{g}|\Delta_R|^2 
    + \int_{{\mathcal{C}}}dt~\frac{1}{2}CV^2,
\end{eqnarray}
where we use shorthand notation $x = ({\bm r},t)$.
Here, $\hat{G}$ denotes the time-ordered Green's function, and $\hat{G}^{-1}$ is a matrix in the left/right space, given by
\begin{eqnarray}
    \label{eq:GLRinv}
    \hat{G}^{-1}(x,x') = \begin{pmatrix}
        \hat{G}_L^{-1}(x)\delta(x-x') & \hat{T}(x,x')\delta(t-t') \\
        \hat{T}^{\dagger}(x,x')\delta(t-t') & \hat{G}_{R}^{-1}(x)\delta(x-x')
    \end{pmatrix}. \nn\\
\end{eqnarray}
The diagonal components, $\hat{G}_L$ and $\hat{G}_R$, represent the on-site propagators of the left and right subsystems, 
and each carries degrees of freedom associated with forward ($+$) and backward ($-$) branches of the Keldysh contour, $\mathcal{C}$.
Their inverse Green's functions are given by
\begin{align}
    \hat{G}^{-1}_{L/R} = \Big[i\hbar\pdv{}{t} - \Big\{-\frac{\hbar^2}{2m}\nabla^2-\mu \Big\}{\sigma_z}\Big]\tau_z \mp \frac{1}{2}e\hat{V}\sigma_z 
    -\hat{\Delta}_{L/R}.  
\end{align}
Here, $\sigma_i$ and $\tau_i$ ($i=x,y,z$) denote the Pauli matrices in Nambu and Keldysh spaces, respectively.
The bosonic fields are represented as matrices in Keldysh space:
\begin{eqnarray}
    \label{eq:V_keldysh}
    \hat{V} &=& \begin{bmatrix}
        V^+ & \\
        & -V^-
    \end{bmatrix}, \\
    \hat{\Delta}_{L} &=& \begin{bmatrix}
        |\Delta^+_{L}|e^{-i\phi_{L}^+\sigma_z}\sigma_x & \\
        & -|\Delta^-_{L}|e^{-i\phi_{L}^-\sigma_z}\sigma_x
    \end{bmatrix},
\end{eqnarray}
and similarly for $\hat{\Delta}_R$.
Here, the complex bosonic field $\Delta_{L}$ is parameterized in terms of amplitude and phase as $\Delta^{\pm}_{L} = |\Delta^{\pm}_{L}|e^{-i\phi^{\pm}_{L}}$.
In what follows, we use the notation $\hat{M}$ to indicate the operator $M$ in Keldysh space, and matrices enclosed by square brackets denote components in the Keldysh basis.

The off-diagonal components in Eq.~\eqref{eq:GLRinv} describe the tunneling processes between the left and right subsystems and are given by
\begin{eqnarray}
    \hat{T} = 
        T_{x,x'}\sigma_z\tau_z.
\end{eqnarray}
Although the structure of $\hat{G}^{-1}$ appears to be diagonal in Keldysh space, this is due to the use of the continuous-time approximation.
It is important to note that $\hat{G}^{-1}$ is properly defined only in conjunction with the full Green's function $\hat{G}$.

To separate the amplitude and phase degrees of freedom of the complex bosonic field, 
we perform a gauge transformation on the Green's function:
$\hat{G}^{-1}\rightarrow \hat{U}\hat{G}^{-1}\hat{U}^{\dagger}$.
The unitary operator $\hat{U}$ is defined as
\begin{eqnarray}
    \hat{U} = \begin{pmatrix}
        \hat{U}_L & \\
        & \hat{U}_R
    \end{pmatrix},~~~
    \hat{U}_L = \begin{bmatrix}
        e^{i\phi_L^+\sigma_z/2} & \\
        & e^{i\phi_L^-\sigma_z/2}
    \end{bmatrix}.
\end{eqnarray}
Under this transformation, the phase dependence is removed from $\hat{\Delta}_{L/R}$,
while the time derivative and spatial derivative are shifted as 
\begin{eqnarray}
    i\hbar \pdv{}{t} &\rightarrow& i\hbar\pdv{}{t} + \frac{1}{2}\hbar\phi^{\pm}_{L/R}\sigma_z, \\
    \nabla &\rightarrow& \nabla - \frac{i}{2}\nabla\phi^{\pm}_{L/R}\sigma_z.
\end{eqnarray}
In addition, the tunneling matrix acquires a coupling to the gauge-invariant phase difference $\phi^{\pm} \equiv \phi_L^{\pm}-\phi_R^{\pm}$ between the two superconductors, and transforms as
\begin{eqnarray}
    \hat{T} \rightarrow \begin{bmatrix}
        T_{x,x'}\sigma_ze^{i\phi^+\sigma_z/2} & \\
        & -T_{x,x'}\sigma_ze^{i\phi^-\sigma_z/2}
    \end{bmatrix}.
\end{eqnarray}

By performing a perturbative expansion in the tunneling matrix, we obtain the effective action in the form
\begin{eqnarray}
    S_{\rm eff} = S_L + S_R + S_{\rm tun}, 
\end{eqnarray}
where $S_L$ and $S_R$ represent the actions of the isolated left and right superconductors, respectively, expressed in terms of $\hat{G}_{L}$ and $\hat{G}_R$.
The tunneling contribution is given by
\begin{align}
    S_{\rm tun} =& i\hbar~\mathrm{Tr} \left[\hat{G}_L \hat{T} \hat{G}_R \hat{T}\right] + \frac{1}{2}i\hbar~\mathrm{Tr} \left[\hat{G}_L\hat{T}\hat{G}_R\hat{T}\hat{G}_L\hat{T}\hat{G}_R\hat{T}\right] \nn \\
    & +\int dt~\frac{1}{2}C({V^+}^2 - {V^-}^2),
\end{align}
where we have expanded the tunneling contribution up to fourth order in the tunneling matrix in order to analyze the Andreev reflection processes.
In the following, we analyze each term separately.

\subsection{Isolated superconductors}
Let us begin by analyzing the action $S_L$ for an isolated superconductor.
The corresponding action for the right superconductor, $S_R$, is analogous, differing only by the sign of the applied bias $V$.
The action $S_L$ is given by
\begin{eqnarray}
    \label{eq:SL}
    S_{L} = -i\hbar~\mathrm{Tr} \ln \hat{G}^{-1}_{L} - \int dx~\frac{1}{g}[|\Delta^+_L|^2 - |\Delta^-_L|^2],
\end{eqnarray}
where
\begin{widetext}
\begin{eqnarray}
    \label{eq:Ginv}
    \hat{G}^{-1}_{L} &=&\Big[ i\hbar\pdv{}{t} - \Big\{-\frac{\hbar^2}{2m}\nabla^2-\mu\Big\}\sigma_z \Big]\tau_z - |\hat{\Delta}_L|\sigma_x + i\hat{\bm{v}}_s\cdot\nabla - \frac{1}{2}m\hat{\bm{v}}_s^2\sigma_z + \frac{1}{2}\Big({\hbar}\pdv{\hat{\phi}_L}{t}- e\hat{V}\Big)\sigma_z 
\end{eqnarray}
and $\bm{v}_s = -(\hbar/2m)\bm{\nabla}\phi_L$.
The hatted bosonic fields, such as $|\hat{\Delta}_L|$, $\hat{\bm{v}}_s$, and $\hat{\phi}_L$, denote matrices in Keldysh space, as defined in Eq.~\eqref{eq:V_keldysh}.
Since we focus on the left electrode, we omit the subscript $L$.
We perform a transformation in Keldysh space---commonly referred to as a ``Wick rotation''---defined by
\begin{eqnarray}
    \hat{G}^{-1}\rightarrow\check{G}^{-1} = \hat{L}\hat{G}^{-1}\tau_z \hat{L}^{\dagger},~~~ \hat{L} = \tau_0 -i\tau_y.
\end{eqnarray}
Hereafter, we use the notation $\check{M}$ to indicate a matrix $M$ in the rotated Keldysh basis.
Along with this transformation, we introduce the classical--quantum (cl--q) representation for the bosonic fields.
As an example, the superconducting amplitude transforms as
\begin{eqnarray}
    |\hat{\Delta}|\rightarrow |\check{\Delta}| = \begin{bmatrix}
        \Delta^{\rm cl} & \Delta^{\rm q}/2 \\
        \Delta^{\rm q}/2 & \Delta^{\rm cl}
    \end{bmatrix},
\end{eqnarray}
where
\begin{eqnarray}
    \Delta^{\rm cl} = (|\Delta^+| + |\Delta^-|)/2,~~~\Delta^{\rm q} = |\Delta^+| - |\Delta^-|.
\end{eqnarray}
Here, $\Delta^{\rm cl}$ and $\Delta^{\rm q}$ are referred to as the classical and quantum components of the field, respectively.
The classical components encode the mean-field dynamics, while the quantum component captures fluctuations around the mean-field solution.

By performing a saddle point approximation with respect to $\Delta^{\rm q}$, one can derive the semiclassical equation of motion for the classical field.
To this end, we expand the action in Eq.~\eqref{eq:SL} up to second order in $\Delta^q$:
\begin{multline}
    S_{L}[\Delta] \approx i\hbar ~\mathrm{Tr} \Bigg[ \check{G}_{\rm BdG}\begin{bmatrix}
        & \Delta^q\sigma_x/2 \\
        \Delta^q\sigma_x/2 & 
    \end{bmatrix}\Bigg] \\
    + \frac{1}{2}i\hbar~\mathrm{Tr} \Bigg[ \check{G}_{\rm BdG}\begin{bmatrix}
        & \Delta^q\sigma_x/2 \\
        \Delta^q\sigma_x/2 & 
    \end{bmatrix}
    \check{G}_{\rm BdG}\begin{bmatrix}
        & \Delta^q\sigma_x/2 \\
        \Delta^q\sigma_x/2 & 
    \end{bmatrix}\Bigg] 
    - \int dx \frac{2}{g}\Delta^c\Delta^q. 
    \label{eq:SL2}
\end{multline}
Here, $\check{G}_{\rm BdG}$ denotes the Green's function incorporating $\Delta^{\rm cl}$ as the anomalous self-energy,
and is expressed in Keldysh-Nambu space as
\begin{eqnarray}
    \check{G}_{\rm BdG} = \begin{bmatrix}
        G_{\rm BdG}^r & G_{\rm BdG}^k \\
        & G_{\rm BdG}^a
    \end{bmatrix},
\end{eqnarray}
with the retarded, advanced, and Keldysh components given by
\begin{eqnarray}
    G_{\rm BdG}^r(\bm{k},\omega) &=& \frac{1}{(\hbar\omega+i\delta)^2-\xi^2(\bm{k})-{\Delta^{\rm cl}}^2}\begin{pmatrix}
        \hbar\omega+i\delta + \xi(\bm{k}) & \Delta^{\rm cl} \\
        \Delta^{\rm cl} & \hbar\omega+i\delta - \xi(\bm{k})
    \end{pmatrix}, \\
    G_{\rm BdG}^a(\bm{k},\omega) &=& \frac{1}{(\hbar\omega-i\delta)^2-\xi^2(\bm{k})-{\Delta^{\rm cl}}^2}\begin{pmatrix}
        \hbar\omega-i\delta + \xi(\bm{k}) & \Delta^{\rm cl} \\
        \Delta^{\rm cl} & \hbar\omega-i\delta - \xi(\bm{k})
    \end{pmatrix}, \\
    \label{eq:Gkel}
    G_{\rm BdG}^k(\bm{k},\omega) &=& [G_{\rm BdG}^r(\bm{k},\omega)-G_{\rm BdG}^a(\bm{k},\omega)]\tanh\left(\frac{\hbar\omega}{2k_{\rm B}\Theta}\right),
\end{eqnarray}
respectively. 
Here, $\Theta$ denotes the temperature and $k_{\rm{B}}$ is the Boltzmann constant. 

By neglecting the second-order term in $\Delta^{\rm q}$ and varying the action with respect to $\Delta^{\rm q}$, we obtain the saddle-point condition:
\begin{eqnarray}
\label{eq:gapeq}
    \frac{\delta S_{L}}{\delta\Delta^{\rm q}}\bigg|_{\Delta^{\rm q} = 0} = 0 ~~~\Rightarrow~~~ \frac{1}{g}\Delta^{\rm cl} = \frac{i\hbar}{4}\tr [G^k_{\rm BdG}\sigma_x],
\end{eqnarray}
which corresponds to the mean-field gap equation for spin-singlet $s$-wave pairing.

Higher-order terms in the quantum component $\Delta^{\rm q}$ describe quantum fluctuations around the mean-field solution.
However, for $s$-wave superconductors, the second-order term with respect to $\Delta^{\rm q}$ in Eq.~\eqref{eq:SL2} identically vanishes.
This implies that amplitude fluctuations of the superconducting order parameter arise only at higher orders and can be safely neglected.
Accordingly, in the following analysis, we treat the superconducting amplitude as a purely classical variable and denote it simply by $\Delta$.

Next, we expand the electromagnetic gauge fields appearing in Eq.~\eqref{eq:Ginv}.
By transforming to the cl--q representation and expanding the action up to the second order, we obtain
\begin{eqnarray}
    \label{eq:Sgauge}
    S_{L}[V,\phi] = \frac{N}{2}\int dx~\Big(\hbar\pdv{\phi^c}{t}-eV^c\Big)\Big(\hbar\pdv{\phi^q}{t}-eV^q\Big) 
    + \int dx~ m\rho_s \bm{v}_s^{\rm cl} \cdot \bm{v}_s^{\rm q},
\end{eqnarray}
where $N$ denotes the density of states at the Fermi level in the normal state.
The quantity $\rho_s$ is the superfluid density, which--at zero temperature--is determined by the diamagnetic response and is given by
\begin{eqnarray}
    \rho_s = \frac{i\hbar}{2}~\mathrm{Tr} [G^k_{\rm BdG}\sigma_z].
\end{eqnarray}
By applying the saddle-point approximation with respect to the classical and quantum components of the scalar potential,
$V^c$ and $V^q$, we obtain the Josephson relations:
\begin{eqnarray}
\label{eq:josephson2}
    \pdv{\phi^{\rm cl}}{t} = \frac{e}{\hbar}V^{\rm cl}, ~~~   \pdv{\phi^{\rm q}}{t} = \frac{e}{\hbar}V^{\rm q},
\end{eqnarray}
which establish the connection between the dynamics of the superconducting phase and the electric voltage potential.

In the presence of a magnetic field, the superfluid velocity is modified via minimal coupling as 
$\bm{v}_s = -(1/2m)(\hbar\bm{\nabla}\phi + 2e\bm{A})$, 
where $\bm{A}$ is the electromagnetic vector potential.
The action associated with the magnetic field is given by
\begin{eqnarray}
    S_{\rm field}[\bm A] = -\frac{1}{4\pi}\int dx~[\bm{\nabla}\times(\bm{A}^{\rm cl} - \bm{A}^{\rm ex})]\cdot(\bm{\nabla}\times\bm{A}^{\rm q}),
\end{eqnarray}
where $\bm{A}^{\rm ex}$ denotes the externally applied vector potential, and $\bm{A}^{\rm cl/q}$ are the classical and quantum components of the vector potential.
By taking the functional derivative of the total action with respect to $\bm{A}^{\rm q}$, we obtain the London equation:
\begin{eqnarray}
    \frac{\delta S}{\delta \bm{A}^{\rm q}} = 0 ~~~ \Rightarrow ~~~ -\rho_se\bm{v}_s^{\rm cl} - \frac{1}{4\pi}\bm{\nabla}\times\bm{\nabla}\times(\bm{A}^{\rm cl} - \bm{A}^{\rm ex}) = 0.
\end{eqnarray}
Consequently, the superconducting current is expressed as
\begin{eqnarray}
    \bm{j}_s = \rho_s\frac{e}{2m}(\hbar\bm{\nabla}\phi^{\rm cl} + 2e\bm{A}^{\rm cl}).
\end{eqnarray}
Note that in Eq.~\eqref{eq:Sgauge}, there are no quadratic terms in the quantum components of the gauge fields.
As a result, fluctuations around these mean-field equations can be safely neglected.

\subsection{Derivation of tunneling action}
In weakly coupled superconducting junctions such as STM, it is reasonable to assume that the superconductors on both sides remain in local equilibrium.
From the analysis in the previous subsections, the superconducting gap is determined by the gap equation \eqref{eq:gapeq},
the voltage potential satisfies the Josephson relations in Eq.~\eqref{eq:josephson2},
and fluctuations in both quantities are negligible.
As a result, the only remaining collective degree of freedom is the superconducting phase difference $\phi=\phi_L - \phi_R$,
which couples to the tunneling matrix element.
By expanding the tunneling action, we now derive the effective action that governs the phase dynamics. 

In the following, we focus on the case of an STM setup, where electron tunneling is assumed to occur locally at a single spatial point.
Due to the uncertainty principle between position and momentum, the momentum of the tunneling electron is not conserved.
As a result, the tunneling matrix element can be taken to be momentum independent, i.e., $T_{kq} = T$.

Under this assumption, the trace over momentum in the tunneling action corresponds to taking the momentum average of each Green's function individually:
\begin{eqnarray}
    \label{eq:Stun1}
    S_{\rm tun} = i\hbar~\mathrm{Tr} \Big[\langle\check{G}_L\rangle_{\bm k}\check{T}\langle\check{G}_R\rangle_{\bm q}\check{T}^\dagger\Big]
    + \frac{1}{2}i\hbar~\mathrm{Tr} \Big[\langle\check{G}_L\rangle_{\bm k}\check{T}\langle\check{G}_R\rangle_{\bm q}\check{T}^\dagger\langle\check{G}_L\rangle_{\bm k}\check{T}\langle\check{G}_R\rangle_{\bm q}\check{T}^\dagger\Big] + \int dt~\frac{\hbar^2C}{4e^2}\Big(\pdv{\phi}{t}\Big)\Big(\pdv{\chi}{t}\Big),
\end{eqnarray}
where $\langle G_{L(R)}\rangle_{\bm k(q)}\equiv\sum_{\bm{k}(\bm{q})}G_{L(R)}(\bm{k}(\bm{q}))$ denotes the momentum-averaged Green's function.
Here, we have applied the Keldysh rotation, and the phase difference is expressed in the cl--q basis as
$\phi = (\phi^+ + \phi^-)/2$ and $\chi = \phi^+-\phi^-$.
The last term in Eq.~\eqref{eq:Stun1} arises from the capacitive energy of junction, obtained by substituting the Josephson relations in Eq.~\eqref{eq:josephson2} into the charging term.
The tunneling matrix in the rotated Keldysh basis takes the form
\begin{eqnarray}
    \check{T} = |T|\begin{bmatrix}
        \sigma_ze^{i\phi\sigma_z/2}\cos\frac{\chi}{4} & e^{i\phi\sigma_z/2}i\sin\frac{\chi}{4} \\
        e^{i\phi\sigma_z/2}i\sin\frac{\chi}{4} & \sigma_ze^{i\phi\sigma_z/2}\cos\frac{\chi}{4}
    \end{bmatrix}.
\end{eqnarray}
On the left side of the junction, we consider an STM tip made of a conventional $s$-wave superconductor.
Its momentum-averaged retarded Green's function is given by
\begin{eqnarray}
    \langle{G}^r_L(\omega)\rangle = \frac{-N_{{L}}\pi}{\sqrt{\Delta^2_{{L}} - (\hbar\omega+i\delta_{{L}})^2}}\begin{pmatrix}
        \hbar\omega+i\delta_{{L}} & \Delta_{{L}} \\
        \Delta_{{L}} & \hbar\omega+i\delta_{{L}} 
    \end{pmatrix} 
    \equiv \begin{pmatrix}
        g_L^r & f_L^r \\
        f_L^r & g_L^r
    \end{pmatrix}.
\end{eqnarray}
The advanced component is given by the {hermitian} conjugate of the retarded one,
and the component can be constructed according to Eq.~\eqref{eq:Gkel}.
On the right side of the junction, we consider a sample of topological superconductors.
The corresponding Green's function is expressed in the spectral representation using the surface DOS $\rho_R(E)$ as
\begin{eqnarray}
    \langle G_R^r(\omega) \rangle = \begin{pmatrix}
        \int dE~\frac{\rho_R(E)}{\hbar\omega-E+i\delta} & f_R^r(\omega)\\
        {f_R^{\dagger r}(\omega)}& \int dE~\frac{\rho_R(E)}{\hbar\omega+E+i\delta}
    \end{pmatrix}
    \equiv \begin{pmatrix}
        g_R^r & f_R^r \\
        f_R^r & \bar{g}_R^r
    \end{pmatrix}.
\end{eqnarray}
In the anomalous components, we denote the off-diagonal part as $f_{R}(\omega)$.
This component typically vanishes in the bulk due to the odd-parity nature of the topological superconductor,
which causes the anomalous component to cancel under momentum averaging.
However, this is not necessarily the case near interfaces, because of the breaking of the translational symmetry. 
Consequently, the ABSs emerge at the interface, which behave as odd-frequency even-parity Cooper pairs distinct from those in the bulk. 
The parity mixing also occurs at interfaces when the superconductor has a large spin-orbit coupling or the interface is magnetically active, which affects the Josephson coupling between even-parity and odd-parity superconductors.
Since the Josephson current is not the main focus of this work, we define $f_R(\omega)$ without specifying its microscopic origin,
and treat the Josephson coupling at an effective level. 
In the main text, we explore the impact of surface ABSs of topological superconductors on the tunneling current and noise.

We now expand the second-order tunneling action, the first term on the right-hand side of Eq.~\eqref{eq:Stun1}.
Since the tunneling matrix is diagonal in Nambu space, the trace can be separated into diagonal and off-diagonal contributions:
\begin{eqnarray}
    S_{\rm tun}^{(2)}=  \frac{1}{2}i\hbar ~\mathrm{Tr} \left[\check{\mathfrak{g}}_{L}\check{T}\check{\mathfrak{g}}_R\check{T}^{\dagger} \right]
    + 
    \frac{1}{2}i\hbar ~\mathrm{Tr} \left[\check{\mathfrak{f}}_{L}\check{T}\check{\mathfrak{f}}_R\check{T}^{\dagger} \right]\equiv S_{\rm tun}^{\mathfrak{gg}} + S_{\rm tun}^{\mathfrak{ff}}.
\end{eqnarray} 
Here, we define $\mathfrak{g}$ and $\mathfrak{f}$ as
\begin{eqnarray}
    \mathfrak{g}^{\eta} = \begin{pmatrix}
        g^{\eta}(\omega) & \\
        & \bar{g}^{\eta}(\omega)
    \end{pmatrix},~~~
    \mathfrak{f}^{\eta} = \begin{pmatrix}
        & f^{\eta}(\omega) \\
         {f^{\dagger}}^{\eta}(\omega) &
    \end{pmatrix},
\end{eqnarray}
where the index $\eta = r, a, k$ labels the retarded, advanced, and Keldysh components in the rotated Keldysh space.
The diagonal term ($S_{\rm tun}^{\mathfrak{gg}}$) corresponds to single-particle tunneling processes, while the off-diagonal part ($S_{\rm tun}^{\mathfrak{ff}}$) describes the Josephson contribution.

We evaluate the diagonal part of the second-order tunneling action:
\begin{eqnarray}
    S_{\rm tun}^{\mathfrak{gg}} &=& 4\hbar\int dtdt'~\bigg[ -\alpha^r(t-t')\sin\frac{\phi(t)-\phi(t')}{2}\sin\frac{\chi(t)}{4}\cos\frac{\chi(t')}{4}\nn \\
    &&+\alpha^a(t-t')\sin\frac{\phi(t)-\phi(t')}{2}\cos\frac{\chi(t)}{4}\sin\frac{\chi(t')}{4}
    + \alpha^k(t-t')\cos\frac{\phi(t)-\phi(t')}{2}\sin\frac{\chi(t)}{4}\sin\frac{\chi(t')}{4}\bigg],
\end{eqnarray}
where $\alpha^{a,r,k}$ denote the self-energies arising from single-particle tunneling processes. 
In the frequency representation, their components are given as
\begin{eqnarray}
    &&\alpha^r(\omega) = \frac{iT^2}{2}\int \frac{d\omega'}{2\pi}~\Big[ g_L^r(\omega+\omega')g_R^k(\omega')  + g_L^k(\omega+\omega')g_R^a(\omega') \Big] \\
    &&\alpha^a(\omega) = \frac{iT^2}{2}\int \frac{d\omega'}{2\pi}~\Big[ g_L^a(\omega+\omega')g_R^k(\omega')  + g_L^k(\omega+\omega')g_R^r(\omega') \Big] \\
    &&\alpha^k(\omega) = \frac{iT^2}{2}\int \frac{d\omega'}{2\pi}~\Big[ g_L^k(\omega+\omega')g_R^k(\omega')  - (g_L^r(\omega+\omega')-g_L^a(\omega+\omega') )(g_R^r(\omega')-g_R^a(\omega')) \Big].
\end{eqnarray}
The retarded and advanced components are related via complex conjugation: $[\alpha^r]^* = \alpha^a$.
The imaginary part, $\alpha^I \equiv \alpha^r-\alpha^a$, describes dissipation in the phase dynamics, which can be expressed as
\begin{eqnarray}
    \alpha^I(\omega) = (iT^2\pi/\hbar) \int dE~\rho_L(E+\hbar\omega)\rho_R(E)\Big[{n_{\rm F}}(E+\hbar\omega)-{n_{\rm F}}(E)\Big]
    \equiv \frac{i}{2e}I_{\rm qp}(\hbar\omega),
\end{eqnarray}
where $I_{\rm qp}(\hbar\omega = eV)$ will later be identified--via the saddle-point equation--as the single-particle current under an applied bias $V$.
This result indicates that incoherent quasiparticle tunneling contributes to the dissipation in the dynamics of the superconducting phase.
While the retarded and advanced components describe the dissipation in the phase dynamics, the Keldysh component encodes the phase fluctuation, which will later be shown to give rise to current noise.
A straightforward calculation yields the Keldysh component as
\begin{eqnarray}
    \alpha^k(\omega) = \alpha^I(\omega)\coth\frac{\hbar\omega}{2k_{\rm B}\Theta},
\end{eqnarray}
which is nothing but the fluctuation--dissipation theorem.

Next, we expand the off-diagonal contribution:
\begin{eqnarray}
    S_{\rm tun}^{\mathfrak{ff}} &=& 4\hbar\int dtdt'~\bigg[ \beta^r(t-t')\sin\frac{\phi(t)+\phi(t')}{2}\sin\frac{\chi(t)}{4}\cos\frac{\chi(t')}{4} \nn\\
    &&+ \beta^a(t-t')\sin\frac{\phi(t)+\phi(t')}{2}\cos\frac{\chi(t)}{4}\sin\frac{\chi(t')}{4} 
    +\beta^k(t-t')\cos\frac{\phi(t)+\phi(t')}{2}\sin\frac{\chi(t)}{4}\sin\frac{\chi(t')}{4}\bigg],
\end{eqnarray}
where the corresponding self-energies are defined in the frequency space as
\begin{eqnarray}
\label{eq:beta_r}
    &&\beta^r(\omega) = \frac{iT^2}{2}\int \frac{d\omega'}{2\pi}~\Big[ f_L^r(\omega+\omega')f_R^k(\omega')  + f_L^k(\omega+\omega')f_R^a(\omega') \Big], \\
\label{eq:beta_a}
    &&\beta^a(\omega) = \frac{iT^2}{2}\int \frac{d\omega'}{2\pi}~\Big[ f_L^a(\omega+\omega')f_R^k(\omega')  + f_L^k(\omega+\omega')f_R^r(\omega') \Big], \\
    &&\beta^k(\omega) = \frac{iT^2}{2}\int \frac{d\omega'}{2\pi}~\Big[ f_L^k(\omega+\omega')f_R^k(\omega')  - (f_L^r(\omega+\omega')-f_L^a(\omega+\omega') )(f_R^r(\omega')-f_R^a(\omega')) \Big].
\end{eqnarray}
As discussed in the main text, in a junction between an $s$-wave superconductor and a topological superconductor, these self-energies typically vanish due to symmetry mismatch between the two sides.
However, this is not necessarily the case when spin-orbit coupling is present, or the junction is magnetically active~\cite{larkin,millis1988quasiclassical}.
Here, we assume that spin-orbit coupling induces an effective $s$-wave component in the topological superconductor.
Under this assumption, we now compute the Josephson effect between two $s$-wave superconductors.

The anomalous Green's functions oscillate at a frequency scale set by the superconducting gap $\Delta$.
As long as we focus on the slow dynamics with characteristic frequency $\omega\ll\Delta$, we can neglect the frequency dependence of the self-energy, an approximation often referred to as the local-time approximation.
Under this assumption, the self-energy components are calculated as
\begin{eqnarray}
\label{eq:Ic}
    &&\beta^r = \beta^a = -T^2\pi^2N_LN_R\Delta\tanh\frac{\Delta}{2k_{\rm B}\Theta} \equiv -I_c/4e, \\
    &&\beta^k = 0,
\end{eqnarray}
where $I_{\rm c}$ is the Josephson critical current originally derived by Ambegaokar, \textit{et. al.}~\cite{ambegaoker1963tunneling}.
As indicated by the vanishing of the Keldysh component, Josephson tunneling is a coherent process and therefore does not involve dissipation or fluctuations.

Next, we consider the fourth-order tunneling contribution to the action.
In the low-bias regime, where the $s$-wave superconductor has no quasiparticle states within the gap, single-particle tunneling is suppressed because the current is proportional to the quasiparticle DOS.
However, in the topological superconductor, gapless quasiparticle states are present.
This enables an Andreev reflection process, in which quasiparticles from the topological superconductor tunnel into the $s$-wave superconductor and form Cooper pairs, and vice versa.

The second term in the right-hand side of Eq.~\eqref{eq:Stun1} represents the Andreev reflection process between the surface state of the topological superconductor and the $s$-wave superconductor.
The action corresponding to this Andreev reflection process originates from the fourth-order tunneling processes, which is obtained from Eq.~\eqref{eq:Stun1} by
\begin{eqnarray}
    S_{\rm tun}^{\mathfrak{fgfg}}=  \frac{1}{2}i\hbar ~\mathrm{Tr} 
    \left[
    \check{\mathfrak{f}}_{L}\check{T}\check{\mathfrak{g}}_R\check{T}^{\dagger} \check{\mathfrak{f}}_{L}\check{T}\check{\mathfrak{g}}_R\check{T}^{\dagger}
    \right].
\end{eqnarray}
Although the calculation is somewhat involved, we now expand the fourth-order tunneling action in powers of the quantum variable $\chi$, retaining terms up to the second order.
The linear term in $\chi$, which captures dissipation due to this tunneling process, is given by 
\begin{align}
    S_{\rm tun}^{\mathfrak{fgfg}}[\chi] =& 4\hbar\int dtdt'dt_1dt_2~  \nn \\
    &\times \bigg\{ \gamma^r(t_1-t,t-t',t'-t_2)\sin\frac{\phi(t)-\phi(t')+\phi(t_1)-\phi(t_2)}{2}\sin\frac{\chi(t)}{4}\cos\frac{\chi(t')}{4}\cos\frac{\chi(t_1)}{4}\cos\frac{\chi(t_2)}{4}, \nn \\
    &- \gamma^a(t_1-t,t-t',t'-t_2)\sin\frac{\phi(t)-\phi(t')+\phi(t_1)-\phi(t_2)}{2}\cos\frac{\chi(t)}{4}\sin\frac{\chi(t')}{4}\cos\frac{\chi(t_1)}{4}\cos\frac{\chi(t_2)}{4} \bigg\},
\end{align}
where the retarded and advanced self-energies are defined as
\begin{eqnarray}
    \gamma^r(\omega_1,\omega_2,\omega_3) &=&  \frac{iT^4}{2}\int \frac{d\omega}{2\pi}~\Big\{f^k(\omega+\omega_1)g^r(\omega+\omega_2)f^r(\omega+\omega_3)g^r(\omega) + f^a(\omega+\omega_1)g^r(\omega+\omega_2)f^r(\omega+\omega_3)g^k(\omega) \nn \\
    && + f^a(\omega+\omega_1)g^r(\omega+\omega_2)f^k(\omega+\omega_3)g^a(\omega)  + f^a(\omega+\omega_1)g^k(\omega+\omega_2)f^a(\omega+\omega_3)g^a(\omega)  \Big\},\\
    \gamma^a(\omega_1,\omega_2,\omega_3) &=& \frac{i T^4}{2}\int \frac{d\omega}{2\pi}~\Big\{ f^k(\omega+\omega_1)g^a(\omega+\omega_2)f^a(\omega+\omega_3)g^a(\omega)  + f^r(\omega+\omega_1)g^a(\omega+\omega_2)f^a(\omega+\omega_3)g^k(\omega) \nn \\
    && + f^r(\omega+\omega_1)g^a(\omega+\omega_2)f^k(\omega+\omega_3)g^r(\omega)  + f^r(\omega+\omega_1)g^k(\omega+\omega_2)f^r(\omega+\omega_3)g^r(\omega) \Big\}.
\end{eqnarray} 
Applying the local-time approximation to the anomalous Green functions of the $s$-wave superconductor allows us to neglect their frequency dependence and to replace them with the averaged frequency $\omega+\omega_2/2$.
Under this assumption, the dissipative part of the self-energy can be evaluated by elementary algebra as 
\begin{eqnarray}
    \gamma^I(\omega_2) &=& \gamma^r(\omega_2)-\gamma^a(\omega_2) \nn \\
    &=& (4\pi^3|T|^4eN_L^2/\hbar)\int_{-\infty}^{\infty}dE~\rho_R(-E+\hbar\omega_2/2)\rho_R(E+\hbar\omega_2/2)\nn\\
    &&\hspace{50pt}\times\Big[~|F_L(E)|\{n_{\rm F}(E-\hbar\omega_2/2)-n_{\rm F}(E) \} +F_L(E)\{n_{\rm F}(E)-n_{\rm F}(E+\hbar\omega_2/2) \}~\Big] \equiv \frac{i}{2e}I_{\rm AR}(\hbar\omega_2/2), \nn \\
\end{eqnarray}
where $I_{\rm AR}(eV)$ denotes the Andreev reflection current.
Similarly, extracting the second-order term of $\chi$ under the local-time approximation yields
\begin{eqnarray}
    S_{\rm tun}^{\mathfrak{fgfg}}[\chi^2] = 2\hbar\int dtdt'~\gamma^k(t-t')\cos(\phi(t)-\phi(t'))\sin\frac{\chi(t)}{2}\sin\frac{\chi(t')}{2}.
\end{eqnarray}
The Keldysh component of self-energy satisfies the fluctuation-dissipation theorem
\begin{eqnarray}
    \gamma^k(\omega) = \gamma^I(\omega)\coth\frac{\hbar\omega}{2k_{\rm B}\Theta}.
\end{eqnarray}

By collecting all contributions, we obtain the total tunneling action in Eq.~\eqref{eq:Action}:
\begin{eqnarray}
        S_{\rm tun} &=& -\frac{\hbar}{2e}\int dt  \bigg[\frac{C}{2e}\pdv[2]{\phi}{t} + I_{\rm c}\sin\phi - {I_{\rm ex}} \bigg]\chi \nn \\
        &-& \frac{\hbar}{2e}\int dtdt'~\bigg[ 2e\alpha^I(t-t')\sin(\frac{\phi-\phi'}{2})\chi - \frac{1}{2}e\alpha^K(t-t')\cos(\frac{\phi-\phi'}{2})\chi\chi' \bigg] \nn \\
        &-&\frac{\hbar}{2e}\int dtdt'~\bigg[2e\gamma^I(t-t')\sin(\phi-\phi')\chi - e\gamma^K(t-t')\cos(\phi-\phi')\chi\chi'\bigg].
        \label{eq:Stunnel_final}
\end{eqnarray}
Here, the quantum variable $\chi$ is assumed to be sufficiently small, and the action has been expanded around $\chi = 0$.
For notational simplicity, we have introduced the shorthand $\phi\equiv\phi(t)$, $\phi'\equiv\phi(t')$, and likewise  $\chi\equiv\chi(t)$, $\chi'\equiv\chi(t')$.

The first line describes the capacitive energy and the Josephson coupling energy,
while the last term in the first line represents the potential energy associated with the external current source $I_{\rm ex}$.
This term arises from the spatial gradient of the superconducting phase induced by the supercurrent, and can be derived by integrating the kinetic energy---given as the second term in Eq.~\eqref{eq:Sgauge}---along the superconducting loop.
The second and third lines account for the effects of single-particle tunneling and the Andreev reflection process, respectively.

\subsection{Current noise via Hubbard-Stratonovich transformation}
In the following, we apply the saddle-point approximation with respect to the quantum variable $\chi$ to derive the semiclassical equation of motion for the classical variable $\phi$. 
But before that,
we first demonstrate how current noise emerges by eliminating the quadratic term in $\chi$.
To eliminate this term, we perform a Hubbard--Stratonovich transformation on the Keldysh components:
\begin{align}
    e^{i\int dtdt'\frac{1}{4}\alpha^k(t-t')\cos(\frac{\phi-\phi'}{2})\chi\chi'} = \int D\xi_1D\xi_2  
    & ~e^{-\int dtdt'\xi_1(t)\frac{1}{4e^2(-i)\alpha^k(t-t')}\xi_1(t')}e^{-\int dtdt'\xi_2(t)\frac{1}{4e^2(-i)\alpha^k(t-t')}\xi_2(t')} \nn \\
    &\times e^{i\int dt \frac{1}{2e}(\xi_1(t)\cos\frac{\phi}{2} + \xi_2(t)\sin\frac{\phi}{2})\chi}
\end{align}
The introduced auxiliary fields $\xi_1$ and $\xi_2$ follow Gaussian statistics with zero mean and correlation:
\begin{eqnarray}
    \langle\xi_\alpha(t)\rangle = 0,~~~\langle\xi_\alpha(t)\xi_{\beta}(t')\rangle=2e^2(-i)\alpha^k(t-t')\delta_{\alpha,\beta},~~~\alpha,\beta = 1,2
\end{eqnarray}
We define the stochastic current due to single-particle tunneling as
\begin{eqnarray}
    I_N^{\rm qp} = \xi_1(t)\cos\frac{\phi}{2} + \xi_2(t)\sin\frac{\phi}{2},
\end{eqnarray}
which obeys the correlation function 
\begin{eqnarray}
    \langle I_N^{\rm qp}(t)I_N^{\rm qp}(t')\rangle = -2ie^2\alpha^k(t-t')\cos\frac{\phi-\phi'}{2}.
\end{eqnarray}
This quantity represents the current noise associated with incoherent single-particle tunneling.

Similarly, for the quadratic term in $\chi$ associated with Andreev reflection, we perform the Hubbard-Stratonovich transformation:
\begin{align}
    e^{i\int dtdt'\frac{1}{2}\gamma^k(t-t')\cos(\phi-\phi')\chi\chi'} = \int D\xi_3D\xi_4 
    &~e^{-\int dtdt'\xi_3(t)\frac{1}{8e^2(-i)\gamma^k(t-t')}\xi_3(t')}e^{-\int dtdt'\xi_4(t)\frac{1}{8e^2(-i)\gamma^k(t-t')}\xi_4(t')} \nn \\
    & \times e^{i\int dt \frac{1}{2e}(\xi_3(t)\cos\phi + \xi_4(t)\sin\phi)\chi}.
\end{align}
\end{widetext}
The corresponding current noise for Andreev reflection processes is then given by the stochastic force
\begin{eqnarray}
    I_N^{\rm AR} = \xi_3(t)\cos{\phi} + \xi_4(t)\sin{\phi},
\end{eqnarray}
which satisfies Gaussian statistics with zero mean and correlation function
\begin{eqnarray}
    \langle I_N^{\rm AR}(t)I_N^{\rm AR}(t')\rangle = -4ie^2\gamma^k(t-t')\cos(\phi-\phi').
\end{eqnarray}

Collecting all contribution involving the phase variable, the total effective action in Eq.~\eqref{eq:Stunnel_final} is recast into
\begin{align}
    S_{\rm tun} =& -\frac{\hbar}{2e}\int dt~\bigg[\frac{C}{2e}\pdv[2]{\phi}{t} + I_{\rm c}\sin\phi - {I_{\rm ex}} + I_N \bigg]\chi \nn \\
    &- \frac{\hbar}{2e}\int dt~\bigg[\int dt'~\bigg\{ 2e\alpha^I(t-t')\sin(\frac{\phi-\phi'}{2}) \nn \\
    &\hspace{45pt}-2e\gamma^I(t-t')\sin(\phi-\phi')\bigg\}\bigg]\chi,
\end{align}
where $I_N(t) =  I_N^{\rm qp}(t) + I_N^{\rm AR}(t) $ denotes the total current noise arising from single-particle tunneling and Andreev reflection processes.
Finally, taking the functional derivative of the action with respect to $\chi$ yields the semiclassical Kirchhoff equation \eqref{eq:RCSJ} for the superconducting circuit, including the current noise.

\section{Fourier representation of Dyson equation\label{Appendix2}}
In this section, we describe the numerical method used to solve the Dyson equation, Eq.~\eqref{eq:Dyson_sigma},
\begin{gather}
    \label{eq:Dyson_sigma2}
    \check{\Sigma} = \check{T}^\dagger + \check{T}^\dagger\circ\check{G}_{RR}^{(0)}\circ \check{T}\circ \check{G}_{LL}^{(0)} \circ \check{\Sigma}.
\end{gather}
To solve this equation, we move to the Fourier representation.
For an isolated system in equilibrium, the Green's function $G^{(0)}(t,t')$ depends only on the time difference $t-t'$ due to the time-translational invariance.
In the present problem, however, the presence of a finite bias voltage breaks this invariance:
the left and right electrodes are described in different rotating frames associated with their respective chemical potentials, and the resulting time-dependent phase factor is incorporated into the left electrode through a gauge transformation~[see Eq.~\eqref{eq:cL_eV}].
As a result, the unperturbed Green’s function of the left electrode takes the form
\begin{align}
    &{G}_{LL}^{(0)\eta}(t,t') = \begin{pmatrix}
        e^{-\frac{i}{\hbar}eV(t-t')}{g}^{\eta}_L(t-t') & e^{-\frac{i}{\hbar}eV(t+t')}{f}^{\eta}_L(t-t') \\
        e^{\frac{i}{\hbar}eV(t+t')}{f}^{\dagger\eta}_L(t-t') & e^{\frac{i}{\hbar}eV(t-t')}{g}^{\dagger\eta}_L(t-t')
    \end{pmatrix} 
\end{align}
where the discrete frequency is defined as $\Omega_n = 2neV/\hbar$.
Here, we use the index $\eta = r, a, k$ to denote the retarded, advanced, and Keldysh components in the rotated Keldysh basis.
We introduce a mixed Fourier transformation of the form
\begin{align}
    G_{LL}^{(0)\eta}(t,t')&= \int\frac{d\omega}{2\pi}\sum_{n}e^{-i\omega(t-t')}e^{-i\Omega_nt}{G}_{LL}^{(0)\eta}(\omega,\Omega_n)
\end{align}
with
\begin{align}
    {G}_{LL}^{(0)\eta}(\omega,\Omega_n)= \begin{pmatrix}
        \delta_{n,0}{g}^{\eta}_{LL}(\omega-eV/\hbar) & \delta_{n,1}{f}^{\eta}_{LL}(\omega+eV/\hbar) \\
        \delta_{n,-1}{f}^{\dagger\eta}_{LL}(\omega-eV/\hbar) & \delta_{n,0}{g}^{\dagger\eta}_{LL}(\omega+eV/\hbar)
    \end{pmatrix} .
\end{align}
The self-energy can be treated in the same manner by performing a mixed Fourier transformation.
Substituting the resulting expressions into Eq.~\eqref{eq:Dyson_sigma2}, we obtain
\begin{widetext}
\begin{eqnarray}
    \check{\Sigma}(\omega,\Omega_n) = \check{T}^{\dagger}\delta_{n,0} + \check{T}^{\dagger}\check{G}_{RR}^{(0)}(\omega+\Omega_n)\check{T}\sum_{n'}\check{G}_{LL}^{(0)}(\omega+\Omega_n',\Omega_n-\Omega_n')\check{\Sigma}(\omega,\Omega_n').
\end{eqnarray}
\end{widetext}
By solving this equation numerically and substituting the result into Eq.~\eqref{eq:Dyson_GLR}, we obtain $G_{LR}$,
from which all other junction Green’s functions can be constructed by substitution into Eqs.~\eqref{eq:GLL}–\eqref{eq:GLR}.

\bibliography{paper}

\end{document}